\documentclass[prd, a4paper, amsfonts, amssymb, amsmath, reprint, showkeys, nofootinbib, twoside]{revtex4-1}
\usepackage[english]{babel}
\usepackage[utf8]{inputenc}
\usepackage[toc,page]{appendix}
\usepackage[colorinlistoftodos, color=green!40, prependcaption]{todonotes}
\usepackage{amsthm}
\usepackage{mathtools}
\usepackage{physics}
\usepackage{xcolor}
\usepackage{graphicx}
\usepackage[left=23mm,right=13mm,top=35mm,columnsep=15pt]{geometry} 
\usepackage{adjustbox}
\usepackage{placeins}
\usepackage[T1]{fontenc}
\usepackage{lipsum}
\usepackage{csquotes}
\usepackage{subfig}
\usepackage{wrapfig}
\usepackage{pinlabel, color}
\usepackage{multirow}
\usepackage{array}

\newcommand{\arccosh}{{\rm arccosh}}
\usepackage[pdftex, pdftitle={Article}, pdfauthor={Author}]{hyperref} 
\usepackage{array}
\newcolumntype{P}[1]{>{\centering\arraybackslash}p{#1}}
\newcolumntype{M}[1]{>{\centering\arraybackslash}m{#1}}
\begin{document}
\title{Holography on tessellations of hyperbolic space}

\author{Muhammad Asaduzzaman, Simon Catterall, Jay Hubisz, Roice Nelson}
\author{Judah Unmuth-Yockey} \email{jfunmuthyockey@gmail.com \\
masaduzz@syr.edu \\
smcatter@syr.edu \\
jhubisz@syr.edu \\
roice3@gmail.com}
\affiliation{Department of Physics, Syracuse University, Syracuse NY 13244}

\date{\today} 

\begin{abstract}
We compute boundary correlation functions for scalar fields on tessellations of two- and three-dimensional hyperbolic geometries.
We present evidence that the continuum  
relation between the scalar bulk mass and the scaling dimension associated with boundary-to-boundary correlation functions survives the truncation of approximating the continuum hyperbolic space with a lattice.
\end{abstract}

\maketitle

\section{Introduction} \label{sec:intro}
The holographic principle posits that the physical content of a gravitational system, with spacetime dimension $d+1$, can be understood entirely in 
terms of a dual quantum field theory living at the $d$-dimensional boundary of that space~\cite{Maldacena:1997re}. This conjecture is not proven, but it is supported by a great deal of
evidence in the case of a gravitational theory in an asymptotically anti-de Sitter space. Furthermore, for
a pure anti-de Sitter space, the dual quantum field theory is conformal. The posited duality can be expressed as an equality between the generating functional for a conformal field theory, and a restricted path integral over fields propagating in AdS:
\begin{equation}
    Z_{\text{CFT}_d} [J(x)] = \int {\mathcal D}\phi~\delta (\phi_0(x) - J(x)) e^{i S_{\text{AdS}_{d+1}}}.
\end{equation}
The boundary values of the fields, $\phi_0$, do not fluctuate, as they are equivalent to classical sources on the CFT side of the duality.

The earliest checks establishing the dictionary for this duality were performed by studying free, massive scalar fields, propagating on pure anti-de Sitter space~\cite{Gubser:1998bc,Witten:1998qj}. These established that the  boundary-boundary two-point correlation function of such fields
has a power law dependence on the boundary separation, where the magnitude of the scaling exponent, $\Delta$, 
is related to the bulk scalar mass, $m_{0}$, via the relation
\begin{equation}
    \label{eq:kw-delta}
    \Delta = \frac{d}{2} \pm \sqrt{\frac{d^2}{4} + m^{2}_{0}},
\end{equation}
where $m_{0}$ is expressed in units of the AdS curvature.  The two choices for the scaling dimension are related to different treatments of the boundary action~\cite{KLEBANOV199989}.  The ``minus'' branch of solutions 
(which can saturate the unitarity bound, $\Delta = d/2-1$) requires tuning to be accessible in the absence of supersymmetry.

In this paper we explore lattice scalar field theory on finite tessellations of negative-curvature spaces to
determine which aspects of the AdS/CFT correspondence survive this truncation.  Finite-volume and discreteness create both ultra-violet (UV) and  
infrared (IR) cutoffs, potentially creating both a gap in the spectrum and
a limited penetration depth from the boundary into the bulk spacetime.  Finite lattice spacing regulates the UV behavior of the correlators. 
Despite these artifacts, we show that such lattice theories do exhibit a sizable regime of scaling behavior, with this ``conformal window'' increasing with total lattice volume.

We specifically construct tessellations of both two- and three-dimensional hyperbolic spaces\footnote{Hyperbolic space is the Euclidean continuation of anti-de Sitter space}, construct scalar lattice actions, and compute the lattice Green's functions to study the boundary-to-boundary correlators. 
We find general agreement with Eq.~\eqref{eq:kw-delta} in the large volume extrapolation.

Prior work has focused on the bulk behavior
of spin models on fixed hyperbolic lattices and on using thermodynamic observables to map the phase diagram \cite{Krcmar_2008,PhysRevE.84.032103,Benedetti_2015,PhysRevE.101.022124}.  Here, the focus is on the structure of the boundary theory
and, since free scalar fields are employed, the matter sector can be computed exactly including the boundary-boundary correlation function.
This setup allows for a direct test of the continuum holographic behavior.

Reference~\cite{brower2019lattice} performed a thorough 
investigation of the scalar field bulk and boundary propagators in two-dimensional hyperbolic space using a triangulated manifold.  Here we 
extend this discussion to other two dimensional tessellations
and to boundary-boundary correlators in three dimensions.

The organization of this paper is as follows.  In Section~\ref{sec:2d} we describe the class of tessellations we use in two
dimensions and the construction of the discrete Laplacian operator needed to study the boundary correlation functions. In Section~\ref{sec:3d} we extend these calculations to three-dimensional
hyperbolic geometry. Finally, we summarise our results in Section~\ref{sec:conclusions}.

\section{Two-dimensional hyperbolic geometry} \label{sec:2d}
Regular tessellations of the two-dimensional hyperbolic plane can be labeled by their Schl\"{a}fli symbol, $\{p,q\}$,
which denotes a tessellation by $p$-gons with the connectivity, $q$, being the number of $p$-gons meeting at a vertex. In order to generate a negative curvature space, the tessellation must satisfy $(p-2)(q-2)>4$. 

We construct our tessellations by first defining the geometry of a \emph{fundamental domain triangle} from its three interior angles. For regular $\{p,q\}$ tessellations, these angles are $\pi/2$, $\pi/p$, and $\pi/q$. In the Poincar\'e disk model, geodesics are circular arcs (or lines) orthogonal to the disk boundary. Circle inversions in the model equate to reflections in the hyperbolic plane and using this property, we recursively reflect this triangle in its geodesic edges to fill the plane with copies of the fundamental region. Each copy corresponds to a symmetry of the tiling. Thus reflections generated by the sides of the triangles form a symmetry group, which is a $(2,p,q)$ triangle group for the $\{p,q\}$ tessellation. The regular tessellation forms from a subset of edges of this group. For a thorough discussion of tessellation construction using the triangle group, see Ref.~\cite{brower2019lattice}.

For the generation of the incidence matrix, it is quite straightforward to build the Laplacian matrix using the connectivity information of the tessellated disk.  In this way, the lattice is stored solely in terms of its adjacency information.  The lattice is then composed of flat equilateral triangles with straight edges, all of which are the same length throughout the lattice.
\begin{figure}
    \centering
    \includegraphics[scale=0.5]{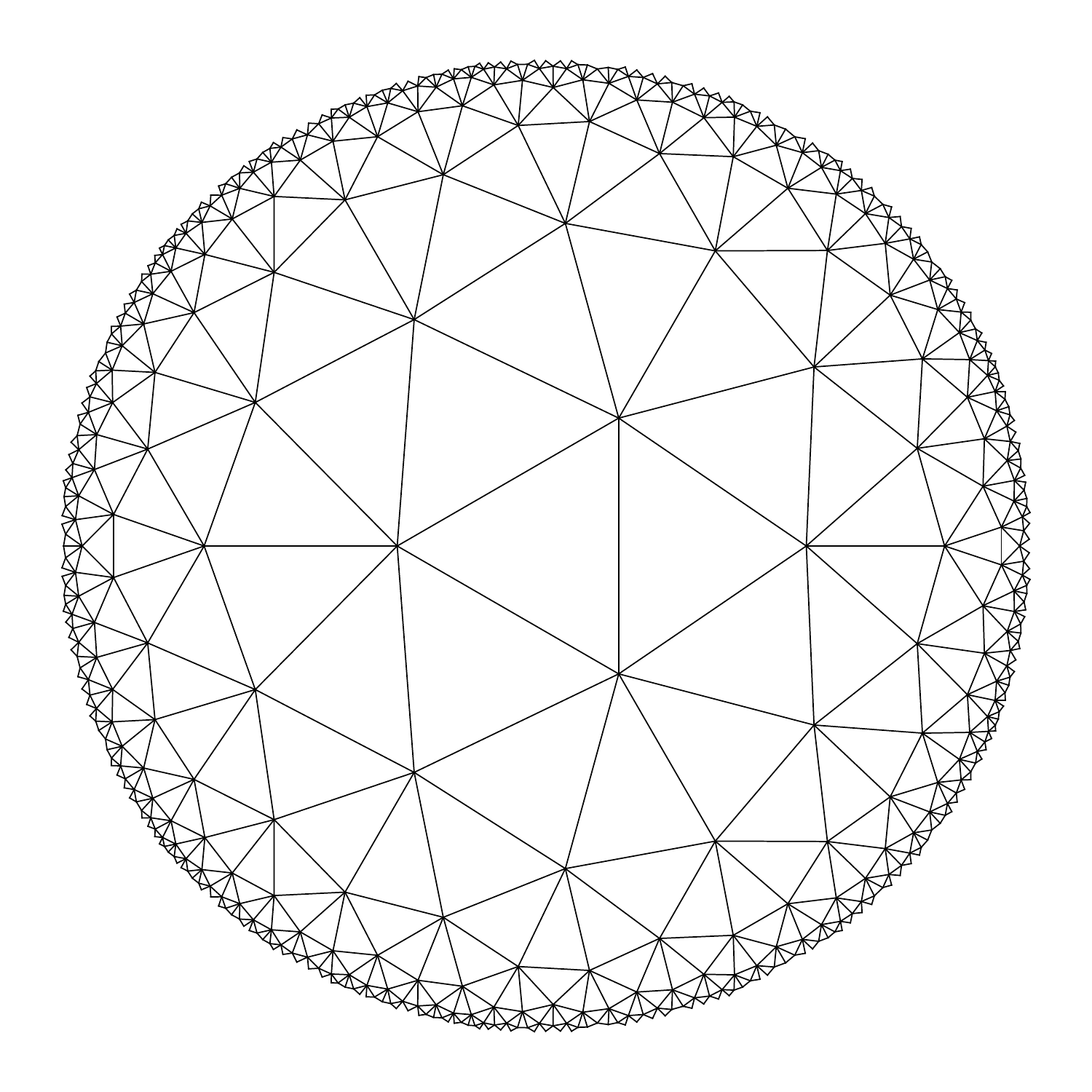}
    \caption{\{3,7\} tessellation in the Poincare Disk model of the hyperbolic space.}
    \label{fig:p3q7}
\end{figure}

\begin{figure}[t]
    \centering
    \includegraphics[width=8.6cm]{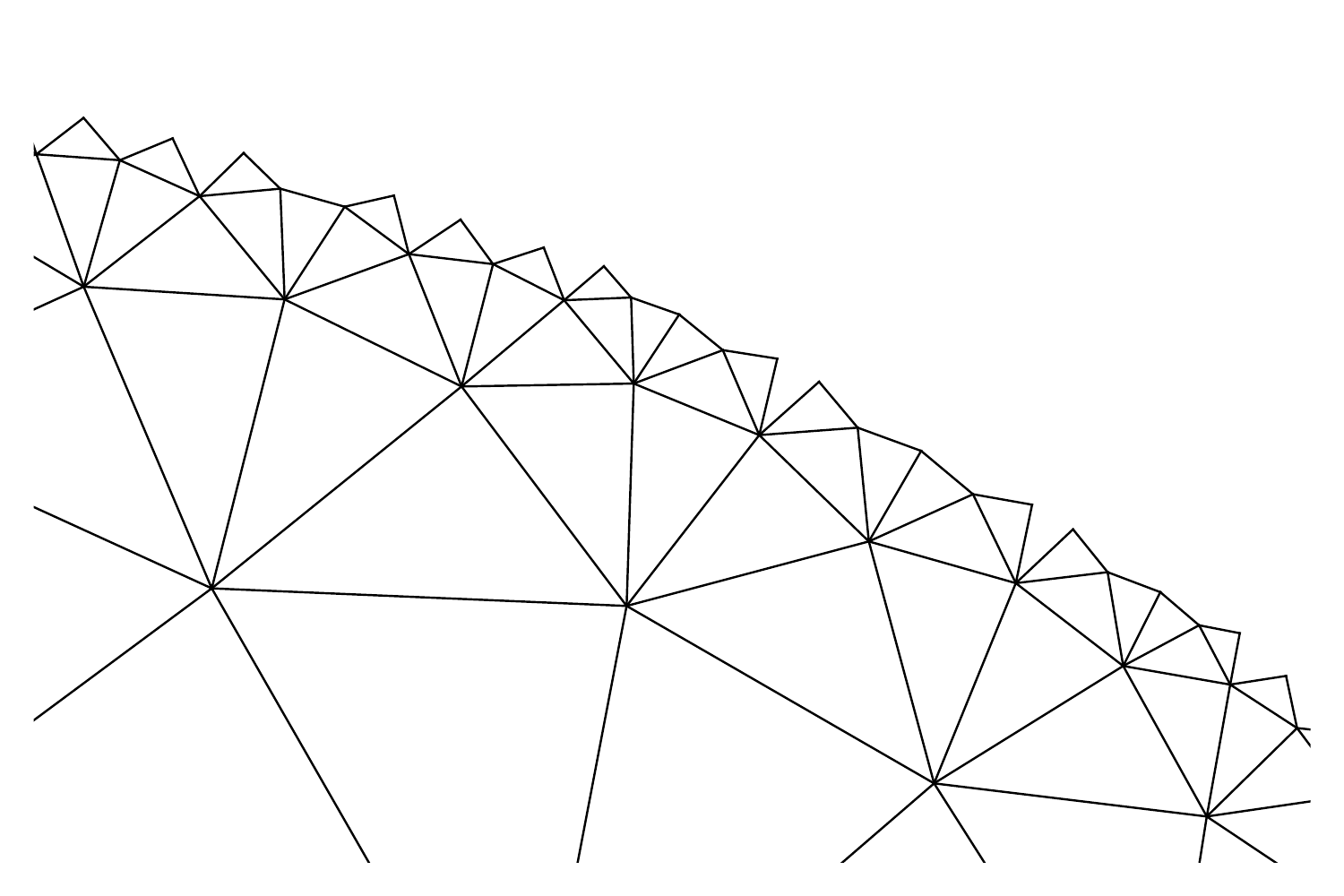}
    \caption{A zoom-in of the boundary of the Poincar\'{e} disk shown in Fig.~\ref{fig:p3q7}.}
    \label{fig:2dbound}
\end{figure}

A typical example of the lattice (projected onto the unit disc) is shown in Fig.~\ref{fig:p3q7}, 
where the tessellation has been mapped onto the Poincar\'{e} disk model and corresponds to the $\{p,q\}$ combination $\{3,7\}$.
An image of the boundary connectivity can be seen in Fig.~\ref{fig:2dbound}.
We see the boundary has all manner of vertex connectivity, some even with seven-fold coordination.  

\begin{figure}[t]
    \centering
    \includegraphics[width=8.6cm]{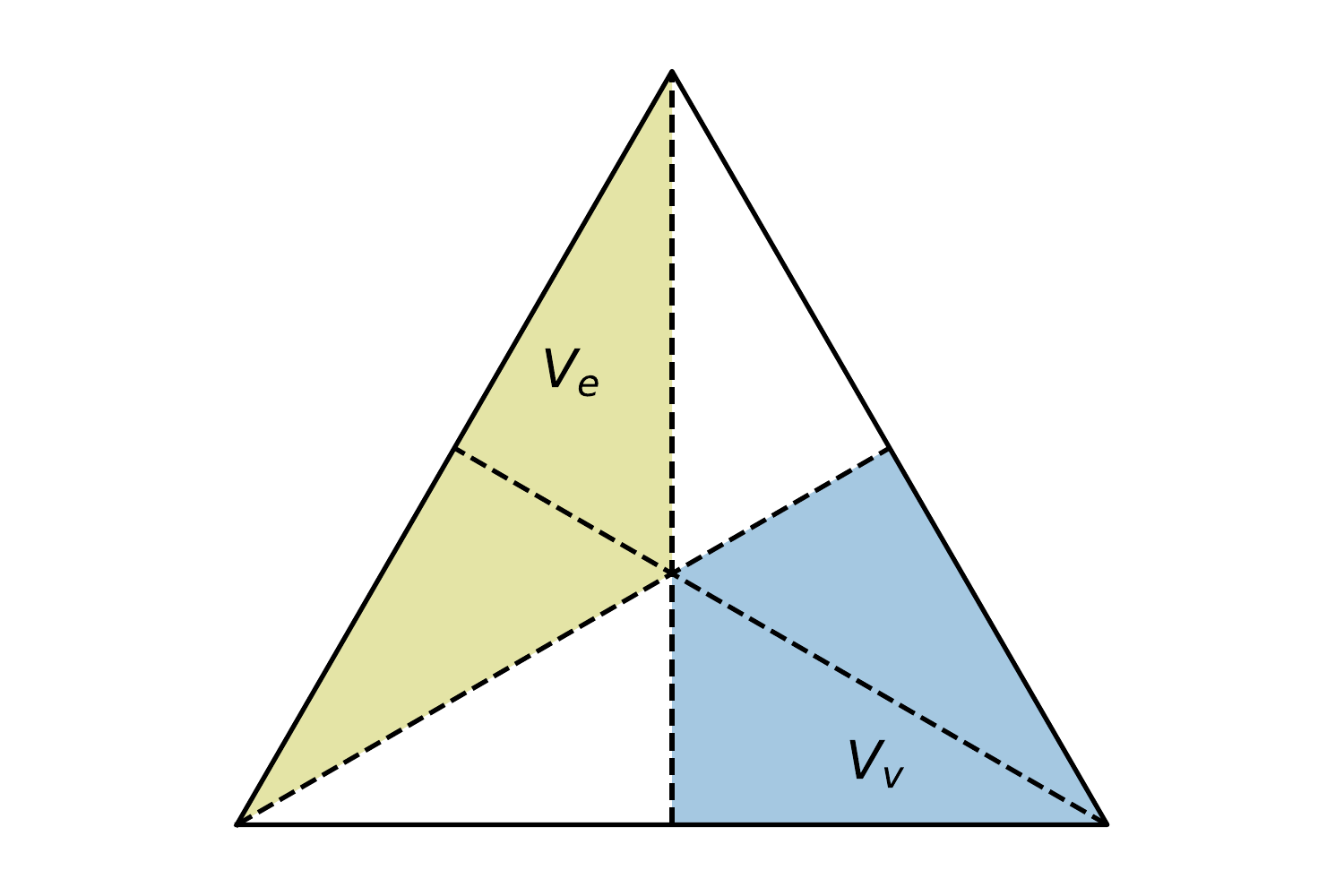}
    \caption{The volume, $V_{e}$, associated with an edge is shown in yellow, and the volume, $V_{v}$, associated with a vertex is shown in blue.  In two dimensions, since each $p$-gon has the same number of vertices as edges, these volumes are always the same (in this case they are both 1/3 of the area of the total triangle).}
    \label{fig:triangle-vols}
\end{figure}

In the continuum, the action for a massive scalar field in two Euclidean spacetime dimensions is given by
    \begin{equation}
        S_{\text{con.}} = \frac{1}{2} \int d^{2}x \sqrt{g} (\partial_{\mu}\phi \, \partial^{\mu} \phi + m^{2}_{0}\phi^{2}),
    \end{equation}
where $m_{0}$ is the bare mass, and $d^{2}x \sqrt{g}$ is the amount of volume associated with each point in spacetime.  The corresponding discrete action on a lattice of $p$-gons is then
\begin{align}
S & =\frac{1}{2}\sum_{\langle x y \rangle} p_{xy} V_e \frac{\left( \phi_x - \phi_y \right)^2}{a^{2}} + \frac{1}{2} \sum_x n_x  m_{0}^2 V_v \phi_x^2 .
\label{eq:2dAction}
\end{align}
Here $V_e$ denotes the volume of the lattice associated with an edge, $V_v$ denotes the volume associated with a vertex, $a$ denotes the lattice spacing, $p_{xy}$ denotes the number of $p$-gons which share an edge (in the case of an infinite lattice this is always two), and $n_x$ is the number of $p$-gons around a vertex.  For two-dimensional $p$-gons,
\begin{equation}
V_v = V_e = \frac{a^2}{4}\cot{\frac{\pi}{p}},
\end{equation}
and an illustration of the volumes associated with links and edges are shown in Fig.~\ref{fig:triangle-vols}.
This definition of the mass and kinetic weights ensures that the sum of the weights gives the total volume of the lattice, 
\emph{i.e.} $\sum_{\langle x y \rangle} p_{xy} V_{e} = \sum_{x} n_{x} V_{v} = A_{p} N_{p}$, where $A_{p}$ is the area of a $p$-gon, and $N_{p}$ is the number of $p$-gons.  
$\sum_{\langle x y \rangle}$ denotes a sum over all nearest-neighbor vertices, and $\sum_{x}$ is over all vertices.  We can write the action from Eq.~\eqref{eq:2dAction} as
\begin{align}
S = \sum_{x, y} \phi_x L_{xy} \phi_y
\end{align}
with $L_{xy}$ given by
\begin{align}
    L_{xy} = -\frac{p_{xy}}{2} \delta_{x, y+\hat{1}} + \frac{1}{2} \left( \sum_{z} p_{xz} \delta_{z, x+\hat{1}} + m_{0}^{2} n_{x} \right)\delta_{x, y}.
\end{align}
In practice we can rescale the scalar field so that the kinetic term has unit weight.

\begin{figure}[t]
    \centering
    \includegraphics[width=8.6cm]{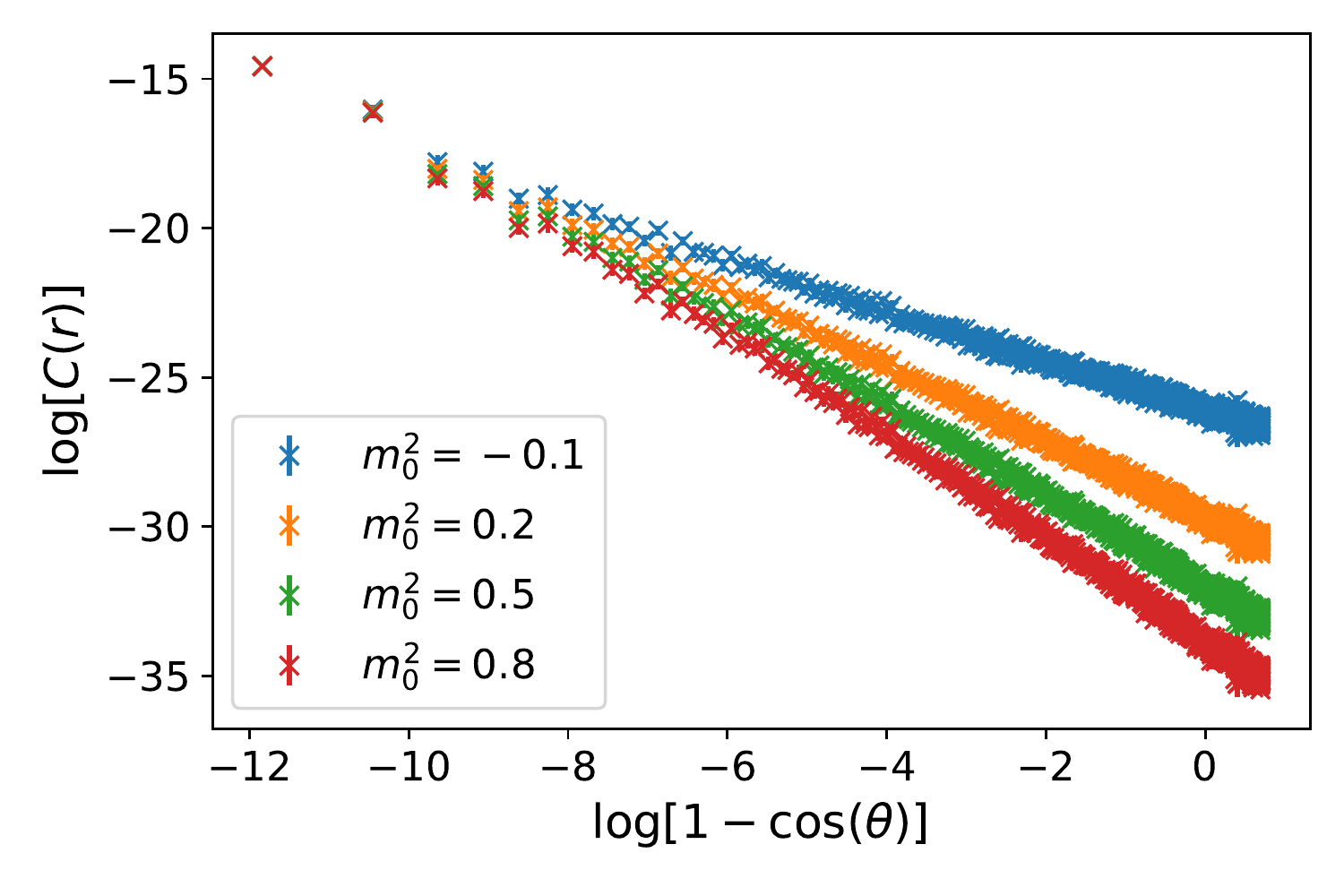}
    \caption{Four different correlators corresponding to different squared bare masses are plotted in log-log coordinates for the case of a 13-layer lattice, with the squared boundary mass set to $M^{2} = 1000$.  The masses here from top to bottom are $m_{0}^{2} = -0.1$, 0.2, 0.5, and 0.8.}
    \label{fig:2d-power}
\end{figure}

We set the edge length, $a$, to one, throughout. In the bulk, for a two-dimensional lattice, $p_{xy} = 2$ and $n_x = q$, but these values change for edges and points on the boundary.  
We note that boundary terms must be added to appropriately approximate the infinite volume AdS/CFT correspondence, in which fields at the AdS boundary are not permitted to fluctuate.  To simulate this, we include a large scalar mass, $M$, only on the boundary vertices and extrapolate fits as $M\to\infty$.
The average boundary correlation function (propagator) is then computed from
\begin{equation}
\label{eq:bb-corr}
C(r) = \frac{\sum_{x,y} L^{-1}_{xy} \delta_{r, d(x,y)}}{\sum_{x, y} \delta_{r,  d(x,y)}},
\end{equation}
where $d(x,y)$ is the distance measured between boundary sites $x$ and $y$ along the boundary.  Looking at this quantity, we observe a power law, $C(r)\sim r^{-2\Delta}$, as can be seen in Fig.~\ref{fig:2d-power},
which shows the correlator for four different masses on a $\{3,7\}$
tessellation with 13 layers containing a total of $N = 2244$ vertices.

To fit the data, we take into account the fact that at finite tessellation depth, the boundary is of finite size which provides an IR cutoff effect.  The modified form of the correlators on the circular compact geometry is known~\cite{Zamolodchikov:2001dz}, with the relation between flat space correlators and those on the circle corresponding to a conformal mapping from the Poincar\'{e} disk to the half-plane, and an additional overall factor associated with the finite size boundary metric~\cite{Brower:2018szu}.  The overall effect is a calculable deviation of the two-point correlation function from straight power law at distances comparable to the perimeter of the circle:
\begin{equation}
    C(r) \propto \frac{1}{\left|1-\cos(\pi \frac{r}{r_\text{max}})\right|^{\Delta}} \underset{r \ll r_\text{max}}{\propto} \frac{1}{| r |^{2\Delta}}
\end{equation}
where $r$ is the distance along the boundary, and $r_\text{max}$ is the distance along the boundary between antipodal points.

We thus fit the correlator using the form,
\begin{align}
    \log C(r) = -\Delta \log \left(1- \cos\left(\theta \right) \right) + k,
\end{align}
with $k$ and $\Delta$ as fit parameters, and $\theta = \pi r / r_\text{max}$.  The error bars are found from using the jackknife method over a subset of the boundary points.  
In addition to the average over boundary points, we also find a non-negligible systematic error from deciding the fit range. 
We calculate this error by repeating the analysis for all different possible reasonable fit ranges and re-sample from these results. 
We add the errors found from this method in quadrature to the jackknife error, and find that the systematic part is by far the largest contribution to the error.

\begin{figure}[t]
    \centering
    \includegraphics[width=8.6cm]{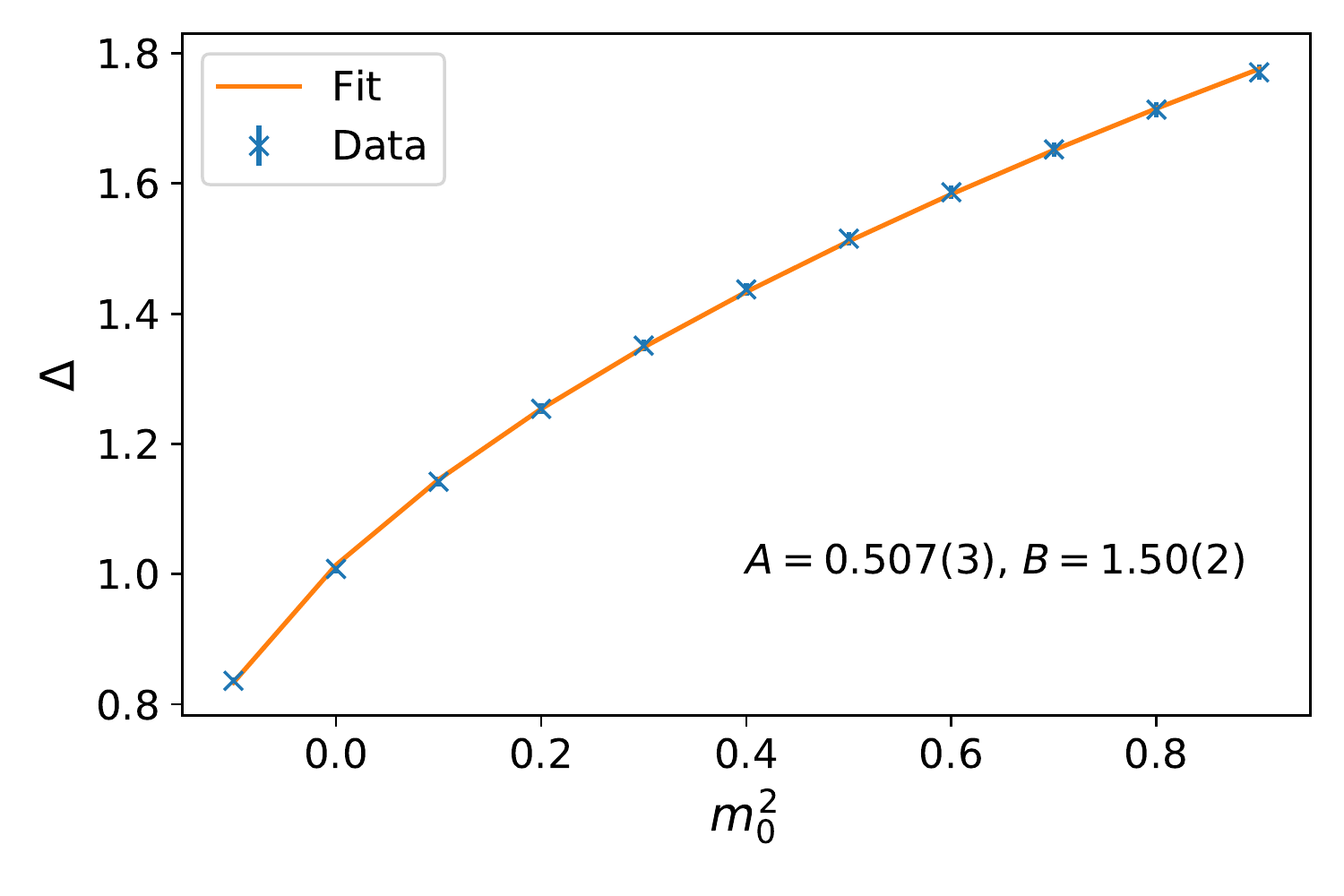}
    \caption{The power law obtained from fitting the correlation function in Eq.~\eqref{eq:bb-corr} versus the squared bare mass.  Here we show the result for a 13-layer lattice with squared boundary mass, $M^2 = 1000$.  We find good qualitative agreement with Eq.~\eqref{eq:kw-delta}.}
    \label{fig:2dscale}
\end{figure}

We check to see if the power, $\Delta$, obeys a similar relation to Eq.~\eqref{eq:kw-delta}, and fit $\Delta$ to the form,
\begin{align}
\label{eq:two-param-fit}
    \Delta = A + \sqrt{A^{2} + B m_{0}^{2}},
\end{align}
where $A$ and $B$ are fit parameters.  
The solid curve in Fig.~\ref{fig:2dscale} indicates the best fit (least squares minimum) to Eq.~\eqref{eq:two-param-fit} for a fixed system size and boundary mass.
We expect $A$ to correspond to half the effective boundary dimension, $d/2$, and $B$ to an effective squared radius of curvature.
We extract the parameters $A$ and $B$ from fits across a range of boundary masses and of system volumes.

Using the various system sizes and boundary masses, we extrapolate the parameters $A$ and $B$ to their values at the infinite-boundary mass, and infinite-system size limit.
In practice, we first extrapolate in system size at fixed boundary mass, and then extrapolate to 
infinite boundary mass at infinite volume. 

\begin{figure}[t]
    \centering
    \includegraphics[width=8.6cm]{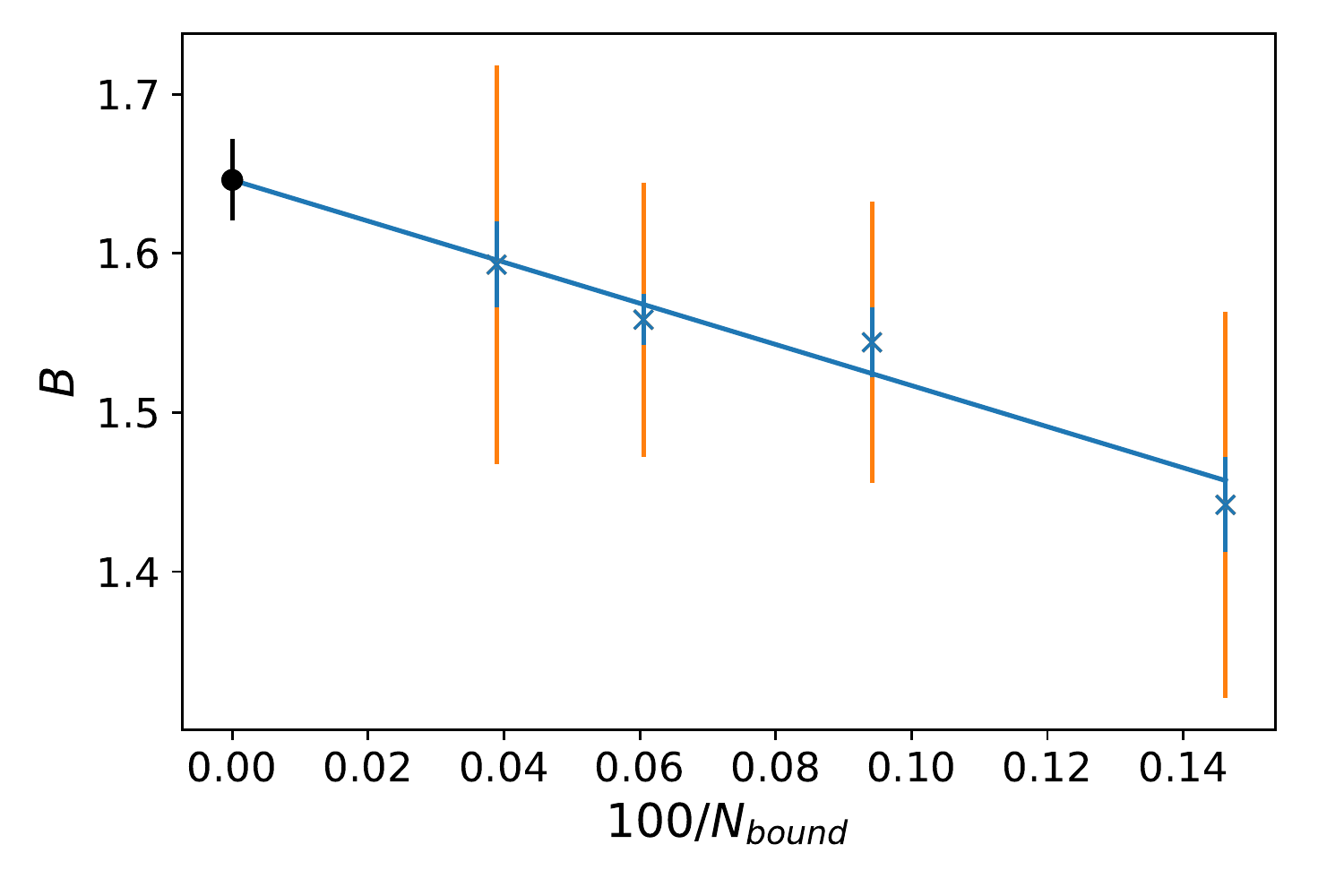}
    \caption{The large-volume extrapolation of the $B$ parameter as a function of the inverse volume of the boundary, for a boundary mass of $M^{2} = 1000$.  We have re-scaled by 100 triangles to remove clutter on the $x$-axis. We fit a line to this data to extract the $B$ parameter at infinite volume, $B_{\infty}$.  We have included the systematic error, due to the choice of the fit range, in orange.}
    \label{fig:2d-bffs}
\end{figure}

In the infinite-volume extrapolation, we identify the regime in which the fit parameters, $A$ and $B$, scale approximately linearly with the inverse boundary size, $N_{bound}$. In other words,
\begin{align}
\label{eq:ffs-eqs}
    A = \frac{C}{N_{bound}} + A_{\infty}, \quad B = \frac{D}{N_{bound}} + B_{\infty},
\end{align}
where $C$, $D$, $A_{\infty}$ and $B_{\infty}$ are fit parameters.  An example of the large-volume data is shown in Fig.~\ref{fig:2d-bffs}.

Once we have $A_{\infty}$ and $B_{\infty}$ for each boundary mass, we extrapolate those values to infinite boundary mass.  Again we look for a window of masses in which the parameters scale  linearly in the inverse squared boundary mass, such that
\begin{align}
    A_{\infty} &= \frac{E}{M^2} + A_{\infty}(M_{\infty}) \\
    B_{\infty} &= \frac{F}{M^2} + B_{\infty}(M_{\infty}),
\end{align}
where $E$, $F$, $A_{\infty}(M_{\infty})$, and $B_{\infty}(M_{\infty})$ are fit parameters.  During our investigation we found we can take the boundary mass sufficiently large such that the extrapolation in large boundary mass is negligible.
We find $A_{\infty} \simeq 0.505(7)$ and $B_{\infty} \simeq 1.65(3)$.
In the appendix we repeat this analysis for the case of $\{4,5\}$ and $\{3,8\}$ tessellations and again find an
effective boundary dimension close to unity but a different value for $B$ reflecting the differing
local curvature.

\section{Three-dimensional hyperbolic geometry} \label{sec:3d}

We now transition to the case of three dimensions.
First we describe the \emph{honeycomb} used in this investigation, as well as its construction. Honeycombs are tilings of three-dimensional space, packings of polyhedra that fill the entire space with no gaps. 

\begin{figure}[t]
    \centering
    \includegraphics[width=8.6cm]{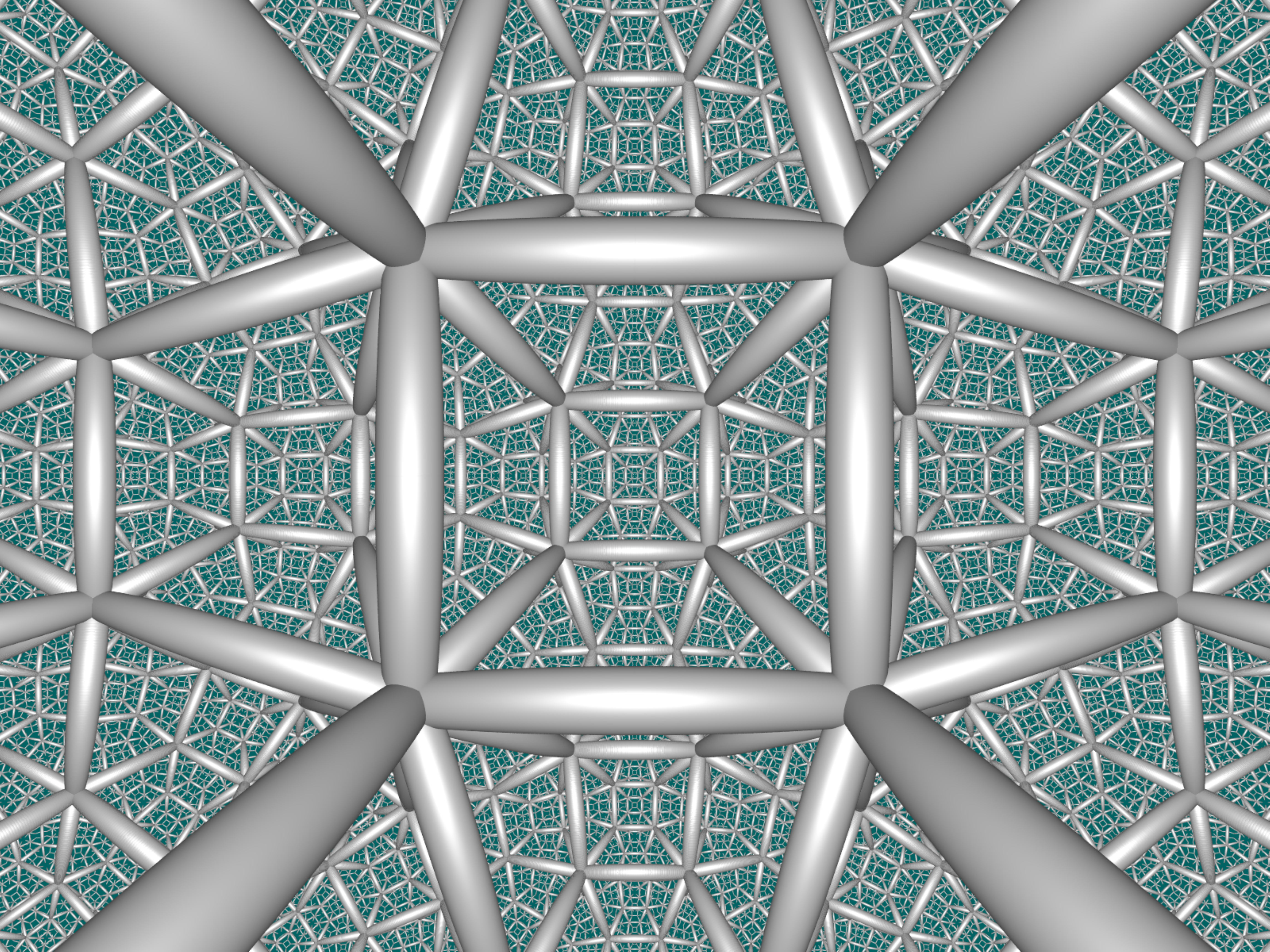}
    \caption{An in-space view of the order-5 cubic honeycomb.}
    \label{fig:o5h}
\end{figure}

Similar to the two-dimensional case, in three dimensions,
one can succinctly describe \emph{regular} honeycombs with a Schl\"afli symbol, a recursive notation for regular tilings. $\{p,q,r\}$ denotes a honeycomb of $\{p,q\}$ \emph{cells}, which are polyhedra (or tilings) of $p$-gons, where $q$ of these surround each vertex~\cite{Coxeter1954}. Here we focus on the $\{4,3,5\}$, also known as the order-5 cubical honeycomb, because the $\{4,3\}$ cubical cells pack $5$ polyhedra around each edge.
A projection of this lattice can be seen in Fig.~\ref{fig:o5h}.
The excess of cubes around an edge gives a local curvature around each edge differing from flat.  Since each Euclidean cube has a face-to-face angle of $\pi/2$, the deficit angle at an edge is $\theta_{d} = 2\pi - 5\pi / 2 = -\pi / 2$. For an infinite lattice, this corresponds to 20 cubes around each vertex, and each vertex has 12 neighboring vertices.

To calculate in hyperbolic space, our code operates natively in the Poincar\'e ball model. In this model, geodesic lines are circular arcs (or lines) orthogonal to the ball boundary, and geodesic surfaces are spheres (or planes) orthogonal to the boundary. Sphere inversions (see~\cite[pp. 124-126]{Needham199902}) in the model equate to reflections in the underlying hyperbolic space and we use this property to build up our honeycomb. 

A general, robust sphere inversion function is key-- one that handles all special cases of spheres or planes reflecting to spheres or planes (as well as the corresponding cases for lines and arcs).  For a thorough discussion of this construction in two dimensions using the triangle group, see Ref.~\cite{brower2019lattice}.

We begin constructing the honeycomb by defining the geometry of a \emph{fundamental tetrahedron} from its six dihedral angles. For regular $\{p,q,r\}$ honeycombs, three of these angles (connected along a zig-zag chain of edges) are $\pi/2$, and the remaining angles are $\pi/p$, $\pi/q$, and $\pi/r$. Using sphere inversion, we recursively reflect the elements of the tetrahedron in its four faces to fill out the space with tetrahedra. Each tetrahedron represents a symmetry of the honeycomb. A fundamental tetrahedron can generate any regular $\{p,q,r\}$ honeycomb with $p,q,r\geq3$~\cite{nelson2017visualizing}.

\begin{figure*}[htbp]
	\centering 
	\subfloat[Central cube.]
	{
		\includegraphics[width=0.24\textwidth]{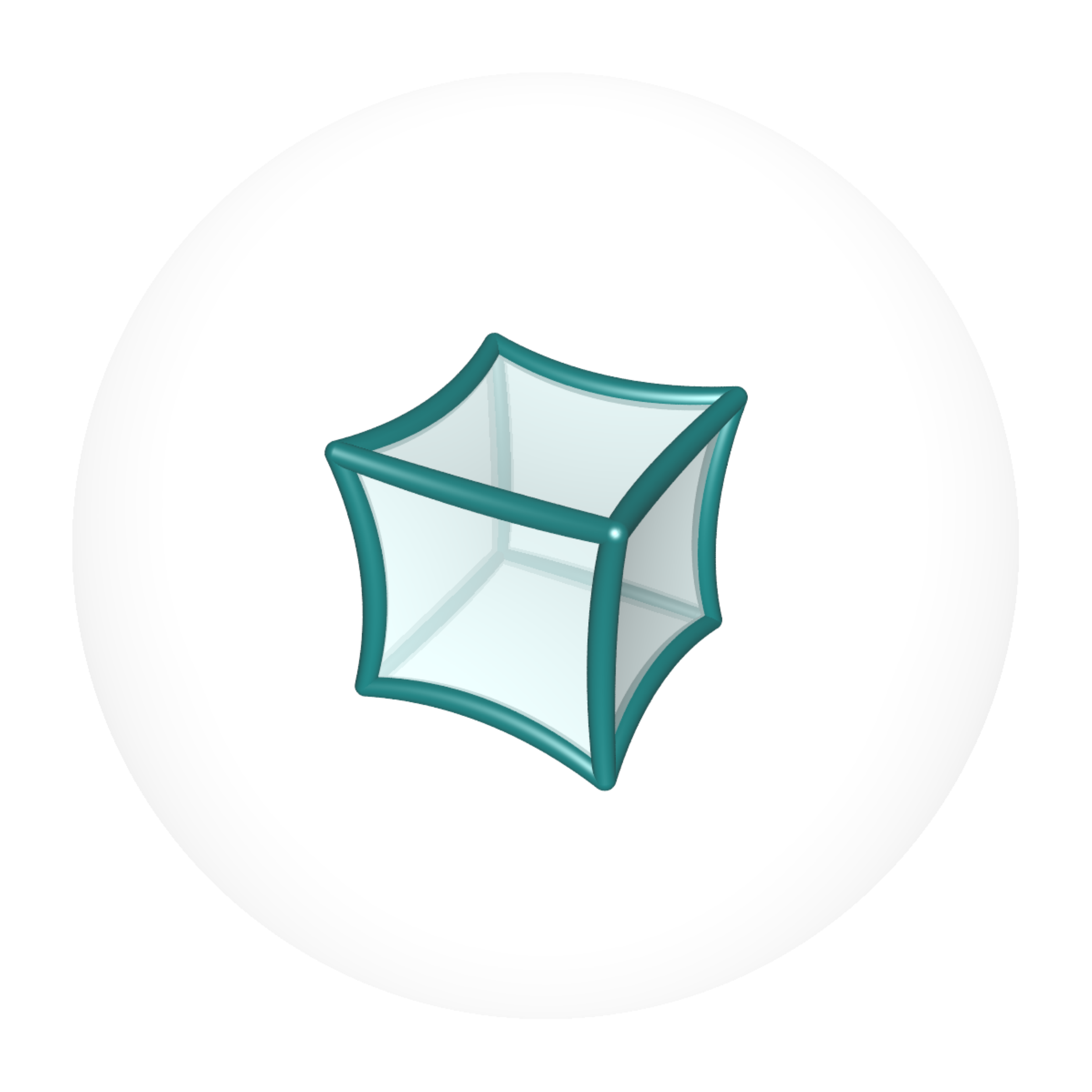}
	}
	\subfloat[Six more.]
	{
		\includegraphics[width=0.24\textwidth]{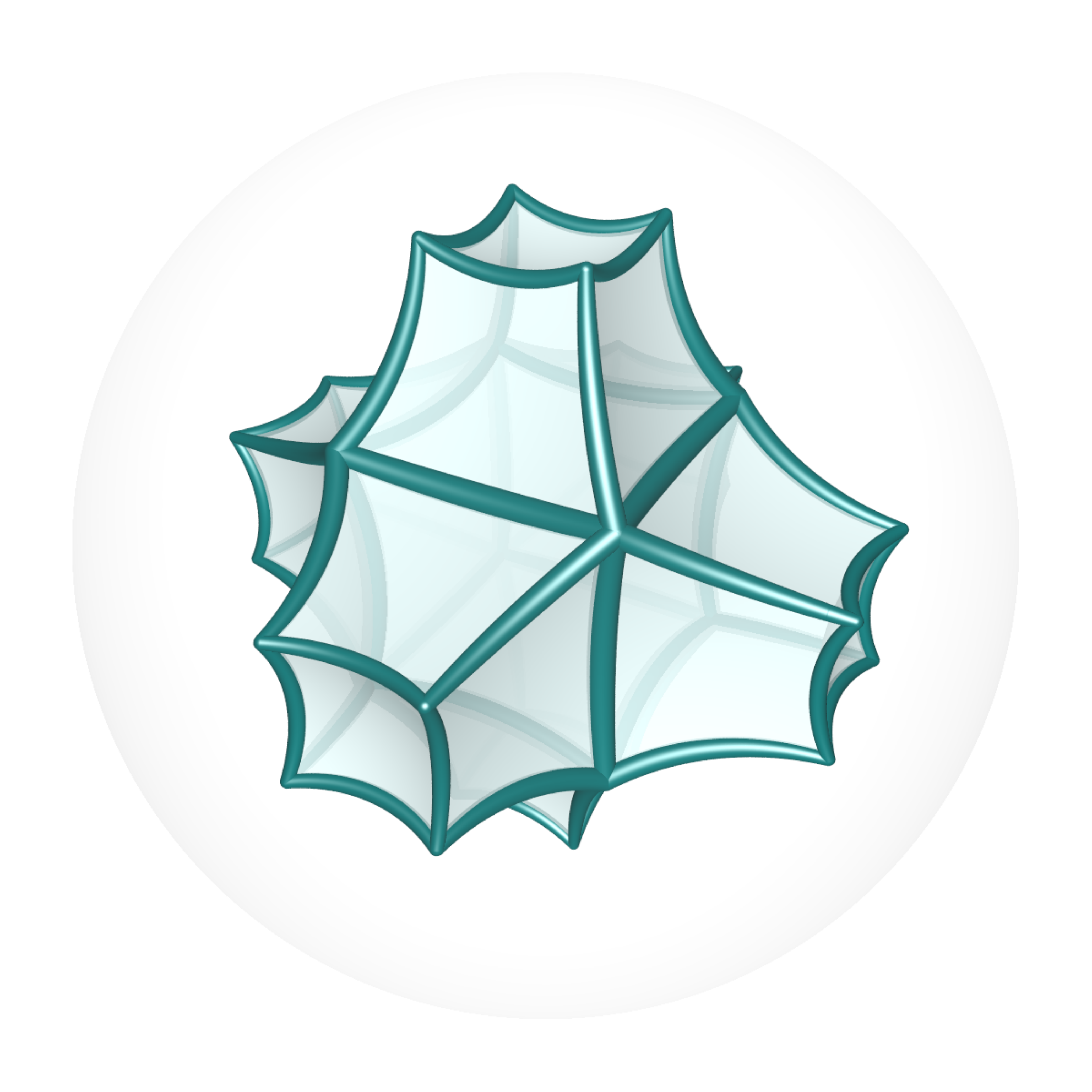}
	}
	\subfloat[Thirty more.]
	{
		\includegraphics[width=0.24\textwidth]{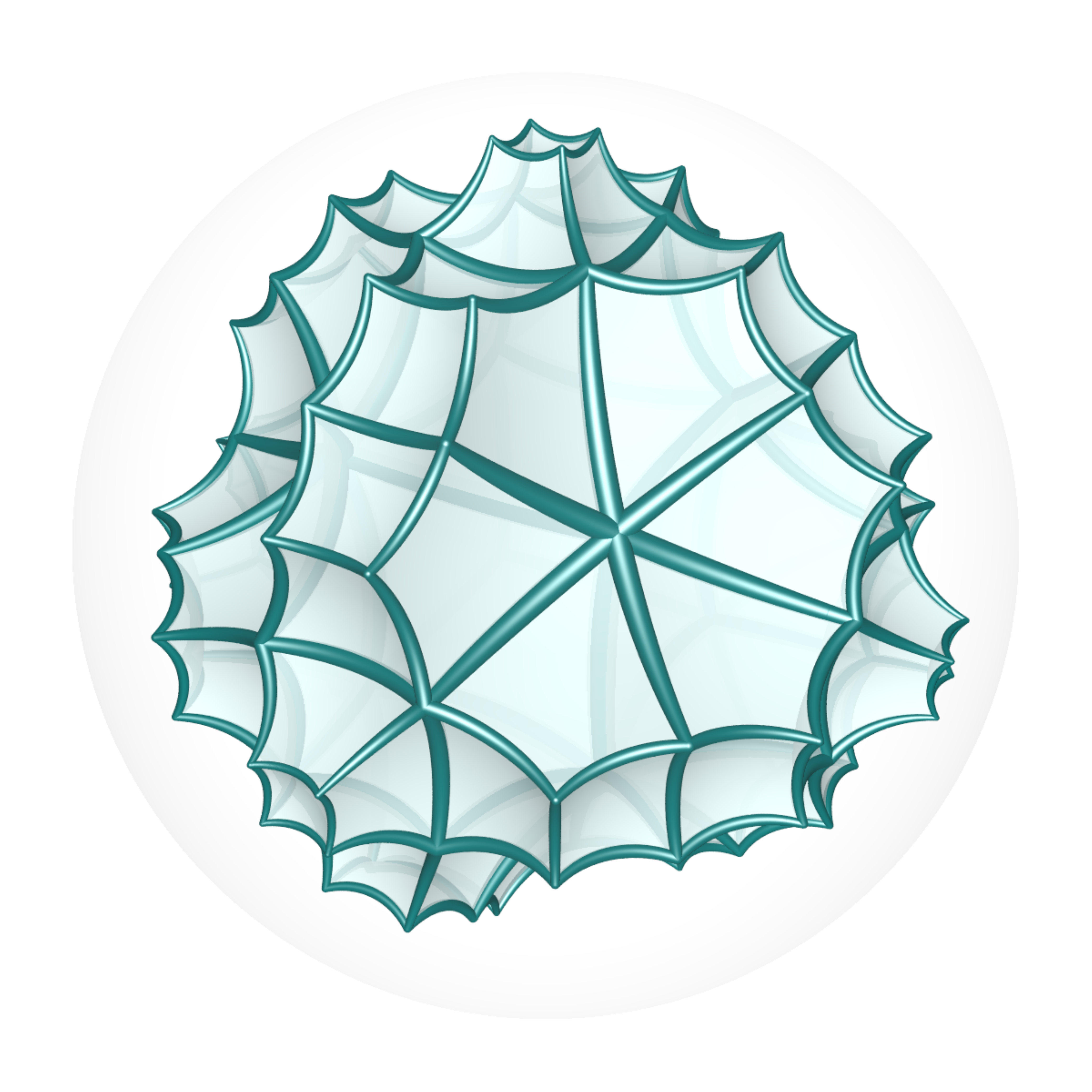}
	}
	\subfloat[126 more.]
	{
		\includegraphics[width=0.24\textwidth]{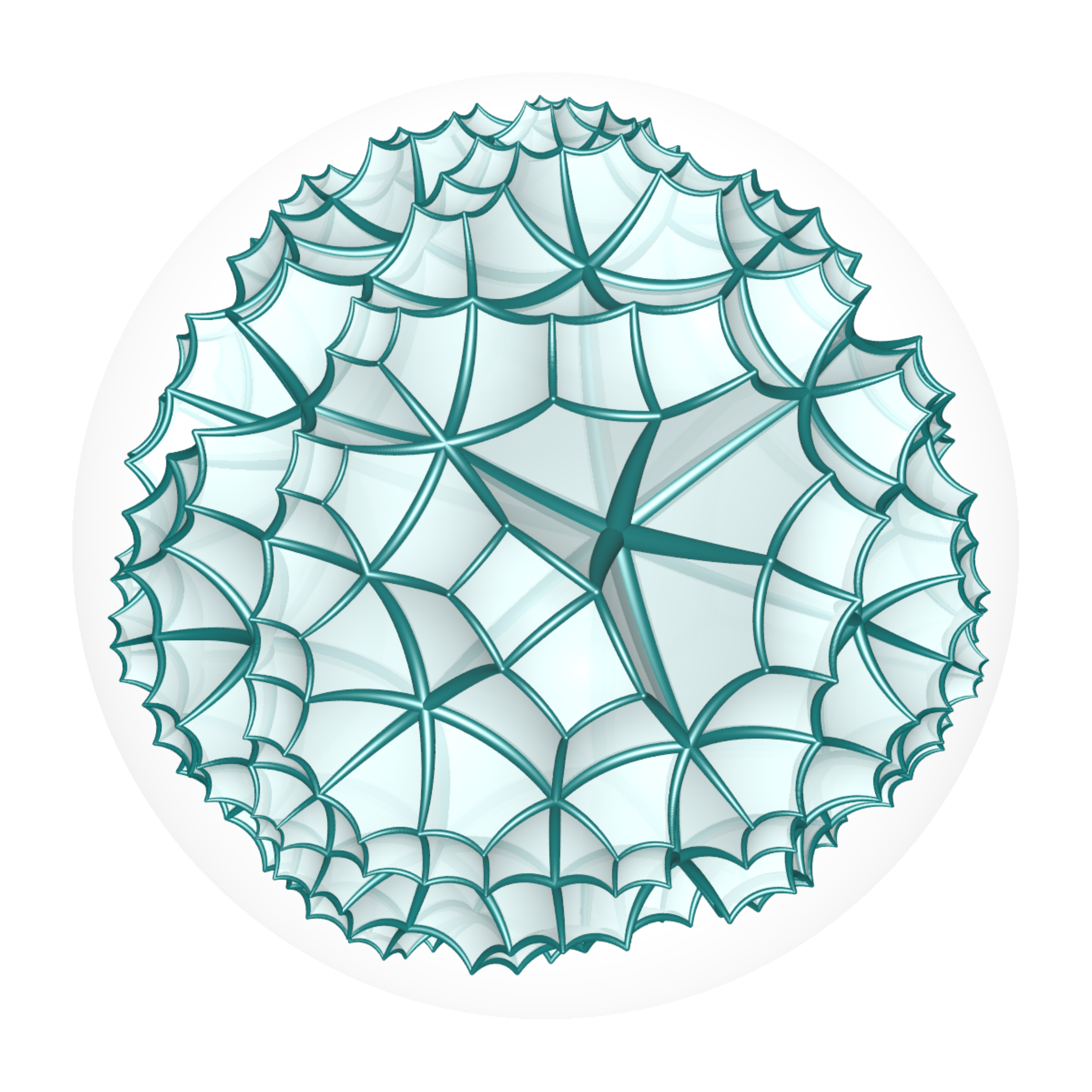}
	}
	\caption{Visualization of step-by-step construction of the lattice with layers of cubes in the $\{4,3,5\}$ honeycomb.  The cube edge lengths appear to vary in length in the Poincar\'{e} ball model; however, the lattice here has a fixed edge length, $a$.}
	\label{fig:435_build_layers}
\end{figure*}

In our case, a set of 48 symmetry tetrahedra form each cube. Reflections in the six faces of these cubes build up the cubical honeycomb in layers of cells, with each successive layer containing all cells one step further in the cell adjacency graph of the honeycomb. The number of cubes in each layer are $1, 6, 30, 126, 498, \dots$, with the total number of cubes up to each level the sum of the entries in this sequence.  This can be seen in Fig.~\ref{fig:435_build_layers}.
We store all the cubes, faces, edges, and vertices that we see during the reflections, taking care to avoid duplicates. In the infinite-volume limit, we would fill the whole of hyperbolic space with cubes.

So far we have described the geometrical construction. We use this information to derive incidence information for all of the elements of the honeycomb, \textit{i.e.}  to determine which vertices, edges, facets, and cubes connect to each other. This we encode as a list of \emph{flags}. A flag is a sequence of elements, each contained in the next, with exactly one element from each dimension. All possible flags encode the full incidence information of our partially-built honeycombs. The incidence encoding is agnostic to geometrical distances. 

We generate lists of flags out to various distances in the cell adjacency graph. The further one recurses, the less edge-effects appear in the incidence information. For example, after adding six layers of cubes, we get enough cells to completely surround all eight vertices of the central cube.

\begin{figure*}[t]
    \centering
    \includegraphics[width=17.2cm]{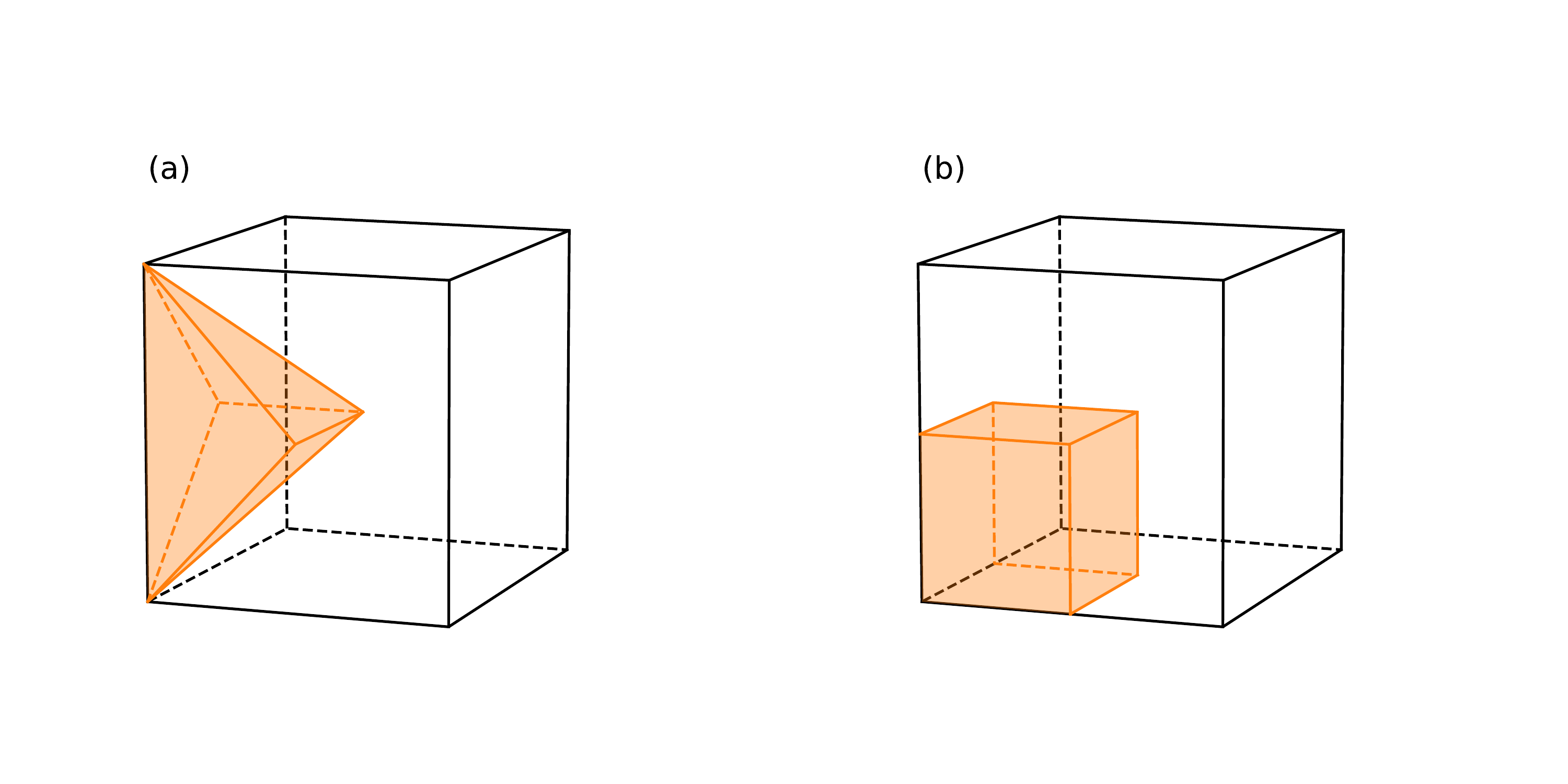}
    \caption{(a) The portion of volume associated with an edge of a cube highlighted, in the text, $V_{e}$. (b) The portion of volume associated with a vertex of a cube highlighted, in the text, $V_{v}$.  For a lattice with uniform edge length $a = 1$, these correspond to $1/12$ and $1/8$ respectively.}
    \label{fig:cube-volumes}
\end{figure*}

Using the lattice described above, we work with the same model as in two dimensions, and take a naive discretization of the scalar field action (this time in three dimensions) given by,
\begin{align}
    S_{\text{lat}} = \frac{1}{2}\sum_{\langle x y \rangle} p_{x y} V_{e} \frac{(\phi_{x} - \phi_{y})^2}{a^2} + \frac{1}{2}\sum_{x} n_{x} V_{v} m^{2}_{0} \phi_{x}^{2}.
\end{align}
Here $\sum_{\langle xy \rangle}$ is over nearest neighbors and $p_{x y}$ is the number of cubes around an edge. In the infinite lattice case, $p_{x y}$ is always five, but we leave it as a variable to allow for consideration of the case when the lattice is finite and has a boundary. $n_{x}$ denotes the number of cubes which share a vertex.  Again, in the infinite case this is always 20, but we leave it as a variable for the finite lattice case. $V_{e}$ and $V_{v}$ are the volumes associated with an edge and a vertex of a cube, respectively.  Since each cube has an edge length of $a$, $V_{e} = a^{3} / 12$ and $V_{v} = a^{3} / 8$.  Illustrations of the two volumes, $V_e$ and $V_v$ are shown in Figs.~\ref{fig:cube-volumes}(a) and~\ref{fig:cube-volumes}(b), respectively.  The weights again are chosen such that $\sum_{\langle x y \rangle} p_{xy} V_{e} = \sum_{x} n_{x} V_{v} = V_{\Box} N_{\Box}$, with $V_{\Box}$ being the volume of a cube, and $N_{\Box}$ the number of cubes.  Above, $a$ is reinserted for clarity but we assume the lattice edge length is one, as before.

We rewrite the lattice action to clearly identify the inverse lattice propagator, even in the presence of a boundary.
To do this, we start by expanding and collecting terms to get
\begin{align}
\nonumber
    S_{\text{lat}} &= \frac{1}{2}\sum_{\langle x y \rangle} p_{x y} V_{e} (\phi_{x} - \phi_{y})^2 + \frac{1}{2}\sum_{x} n_{x} V_{v} m^{2}_{0} \phi_{x}^{2} \\ \nonumber
    &= -\frac{1}{2}\sum_{\langle x y \rangle} p_{x y} V_{e} (\phi_{x} \phi_{y} + \phi_{y} \phi_{x}) \\
    &+ \frac{1}{2} \sum_{x} \left(\sum_{y} p_{x y}V_{e} + m_{0}^{2} n_{x} V_{v} \right) \phi_{x}^{2},
\end{align}
where $\sum_{y}$ in the second term is over points neighboring vertex $x$.  Using the fact that $V_{e} = (2/3)V_{v}$, we simplify further to get
\begin{align}
\nonumber
    S_{\text{lat}} = &-\sum_{\langle x y \rangle} \frac{p_{x y}}{3} (\phi_{x} \phi_{y} + \phi_{y} \phi_{x}) \\
    &+ \sum_{x} \left(\sum_{y} \frac{p_{x y}}{3} + \frac{m_{0}^{2}}{2} n_{x} \right) \phi_{x}^{2}.
\end{align}
We express the action in terms of the inverse lattice propagator and get
\begin{align}
\nonumber
    S_{\text{lat}} &= S_{\text{kinetic}} + S_{\text{mass}} \\
    &= \sum_{x,y} \phi_{x} L_{x y} \phi_{y}
\end{align}
with
\begin{align}
\label{eq:lapkin}
    S_{\text{kinetic}} = -\sum_{x, y} \phi_{x} \frac{p_{xy}}{3} \delta_{x, y + \hat{1}} \phi_{y}
\end{align}
and
\begin{align}
\label{eq:masslap}
    S_{\text{mass}} = \sum_{x, y} \phi_{x} \left( \left( \sum_{z} \frac{p_{xz}}{3} \delta_{z, x+\hat{1}} \right) + \frac{m_{0}^{2}}{2} n_{x} \right)\delta_{x, y} \, \phi_{y}.
\end{align}
In the case of an infinite lattice, this simplifies to
\begin{align}
    L_{xy} = -\delta_{x, y+\hat{1}} + 12 \left( 1 + \frac{m_{0}^{2}}{2} \right) \delta_{x, y},
\end{align}
which is expected for a lattice with 12-fold coordination.
Using Eqs.~\eqref{eq:lapkin} and~\eqref{eq:masslap}, we construct an inverse lattice propagator for the hyperbolic lattice considered here, and use it in numerical computations.

\begin{figure}
    \centering
    \includegraphics[width=8.6cm]{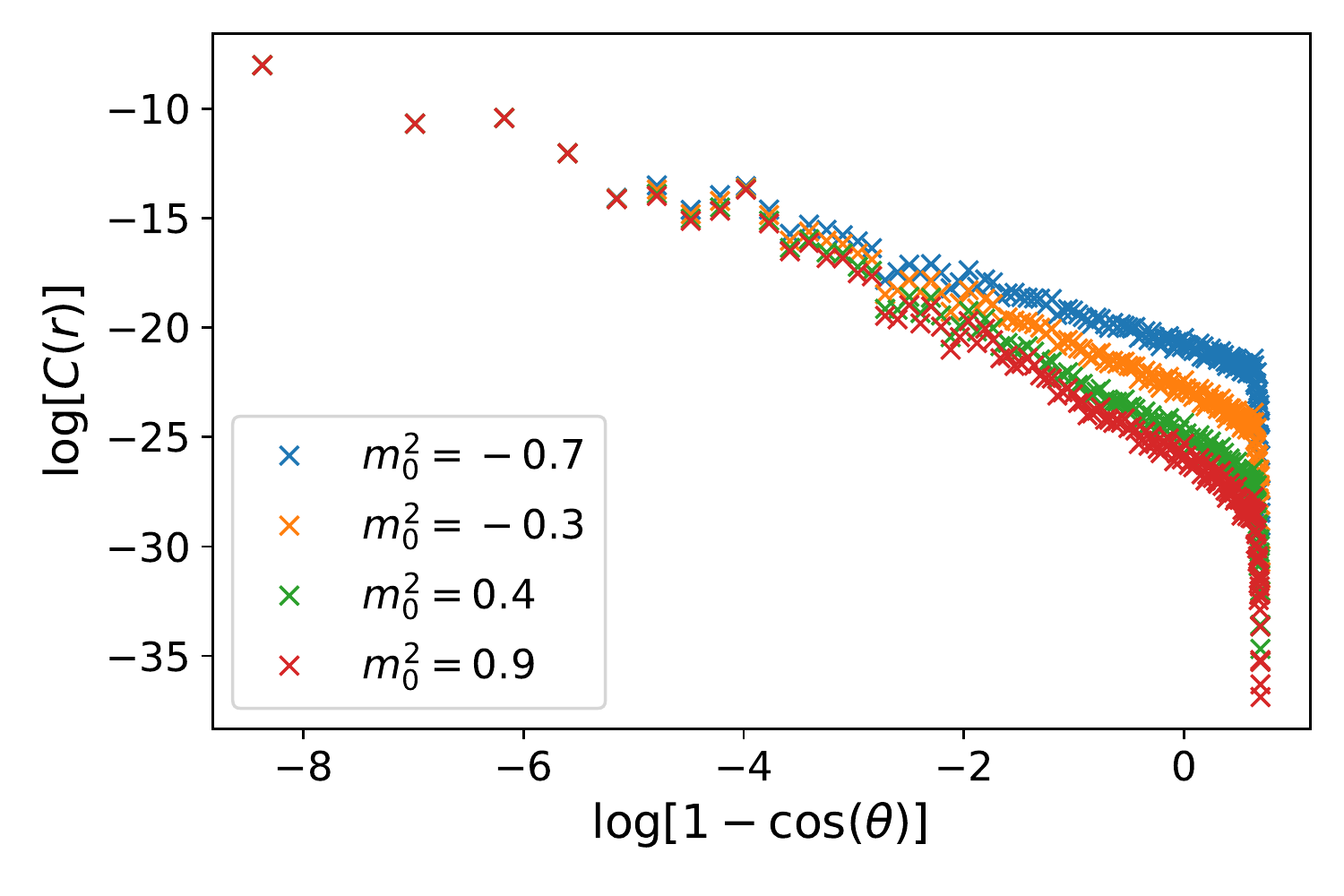}
    \caption{A log-log plot of the boundary-boundary correlator, $C(r)$, as a function of the angle between source and sink along the boundary, for a seven-layer lattice.  Here the bulk masses from top to bottom are $m^{2}_{0} = -0.7, -0.3, 0.4$, and $0.9$, and the boundary mass is $M^2 = 10$.}
    \label{fig:corr3d}
\end{figure}

Again, the boundary correlator is given by inverting the matrix corresponding to the discrete scalar inverse propagator. A typical set of correlators are shown in
Fig.~\ref{fig:corr3d}, corresponding to four bulk masses and squared boundary mass $M^2 = 10$.
We take multiple sources on the boundary and compute the one-to-all correlator for those sources on the boundary. In Fig.~\ref{fig:corr3d} we again plot the correlator as a function of
$1-\cos{\left(\pi r / r_{\text{max}} \right)}$.  The geodesic distance $r$ is computed by starting at the source vertex, taking a step to all neighboring boundary vertices, then taking a step from those vertices to their neighboring boundary vertices, skipping vertices that have already been visited, and so on, until all vertices have been visited.  The error bars are produced using the jackknife method on the sources.

Clearly, a distance window exists in which the correlator follows
a power law.  This power-law behavior is observed for all masses explored in this study, and seems to solely be a consequence of the lattice geometry.  We fit a power law to this window for a series of fixed, squared bulk mass.  By far the largest source of error in this analysis is the systematic error in choosing a fit range.  To improve this error, we bin the data in the regime of interest, and vary the fit range in the binned data.  By resampling from all the reasonable fit ranges we acquire a systematic error.
The jackknife error from the sources is added in quadrature with this systematic error to produce the final errors on each power-law fit.
From this fit, we obtain the power versus the squared mass.

\begin{figure}[t]
    \centering
    \includegraphics[width=8.6cm]{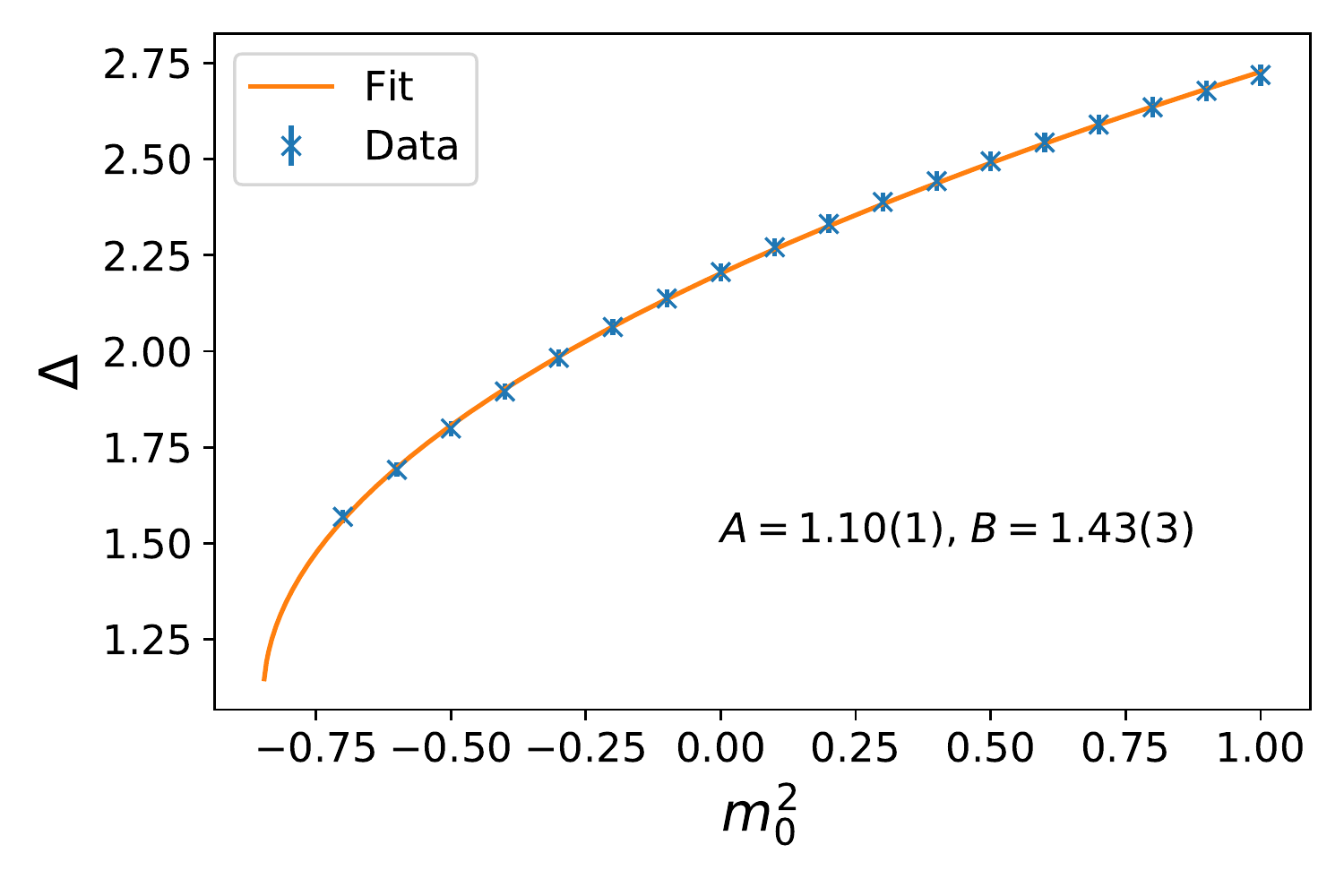}
    \caption{A fit of $\Delta$ versus the squared mass using Eq.~\eqref{eq:two-param-fit} for a six-layer lattice.  Here the boundary mass is $M^{2} = 10$.}
    \label{fig:6layer-2param}
\end{figure}

In the continuum, in the case of anti-de Sitter space, the boundary-boundary two-point correlator is expected to show the behavior from Eq.~\eqref{eq:kw-delta} with boundary dimension $d = 2$.  We attempt a fit using Eq.~\eqref{eq:two-param-fit}.
An example of the fits can be seen in Fig.~\ref{fig:6layer-2param}.
We note that the power, $\Delta$, is well-defined even in the regime of negative squared mass, indicating the operator $-\nabla^{2} + m_{0}^{2}$ is positive in this regime.  Based on these numerical results, the behavior of $\Delta$ here matches well with the expected behavior of $\Delta_{+}$ expressed in Ref.~\cite{KLEBANOV199989}. 

\begin{figure}[t]
    \centering
    \includegraphics[width=8.6cm]{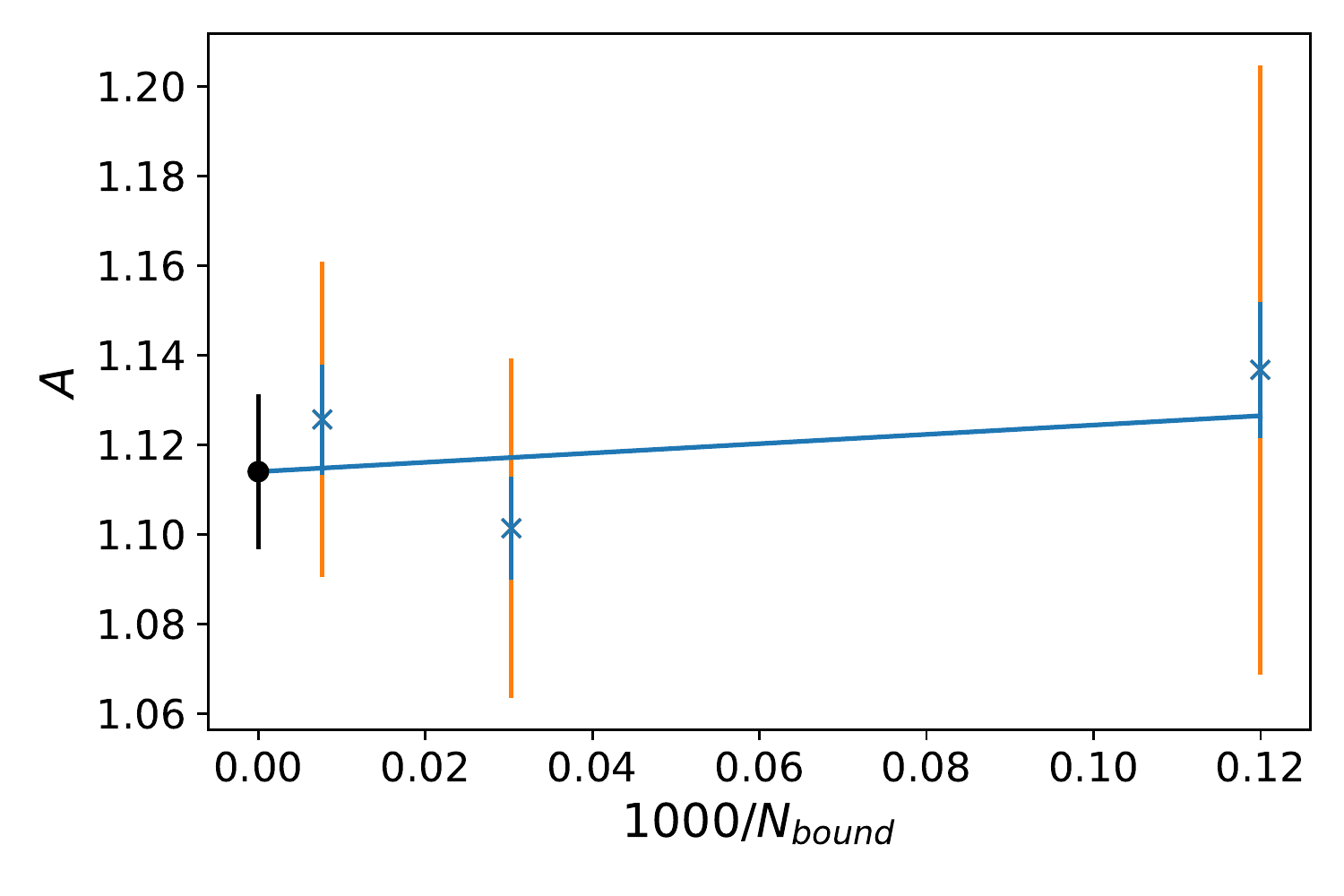}
    \caption{The finite-size scaling of the fit parameter, $A$, from Eq.~\eqref{eq:two-param-fit}.  The volumes have been rescaled by 1000 for readability.  All three volumes use the same boundary mass of $M^{2} = 10$.}
    \label{fig:3d-ffs-a}
\end{figure}

\begin{figure}[t]
    \centering
    \includegraphics[width=8.6cm]{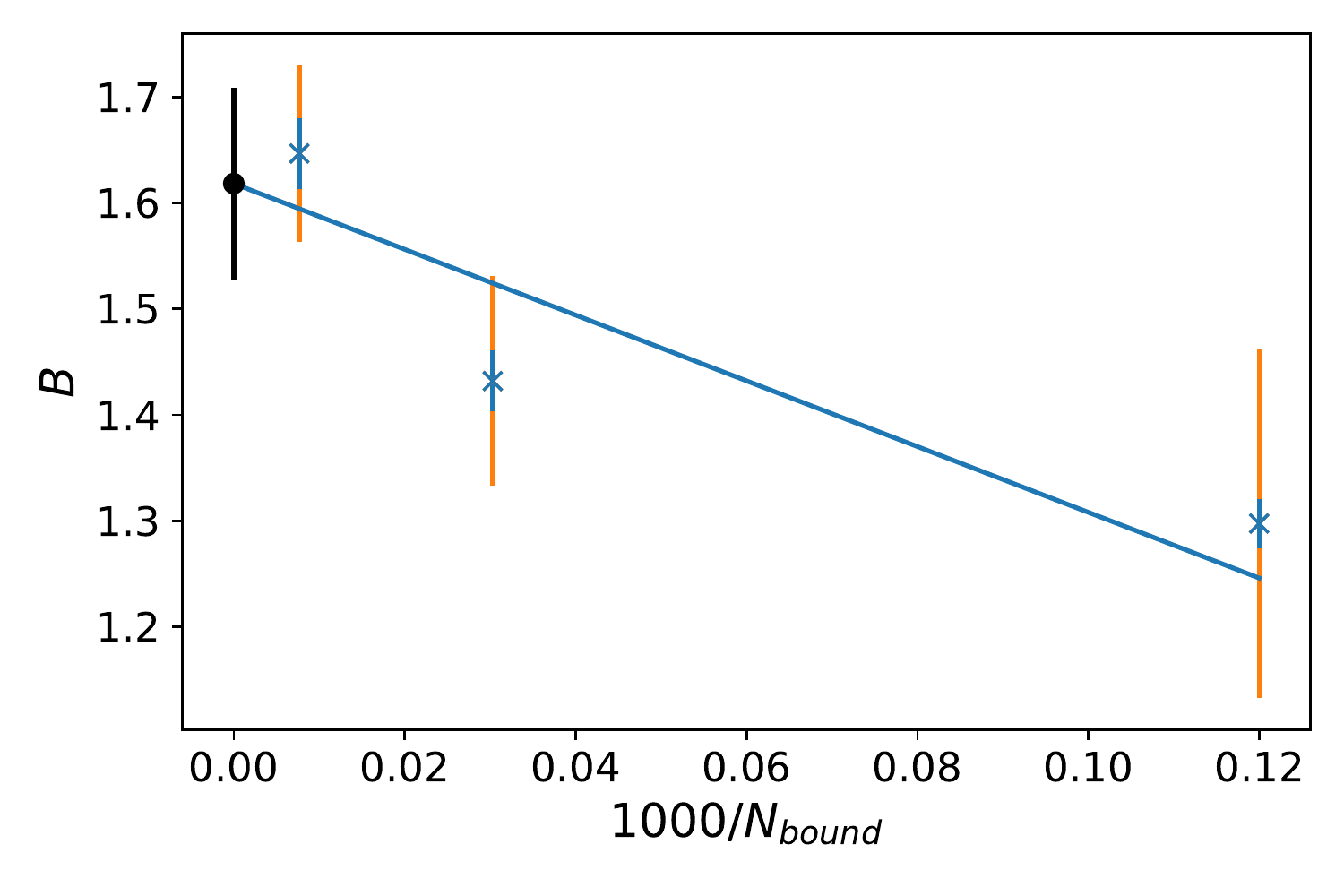}
    \caption{The finite-size scaling of the fit parameter, $B$, from Eq.~\eqref{eq:two-param-fit}.  The volumes have been rescaled by 1000 for readability.  All three volumes use the same boundary mass of $M^{2} = 10$.}
    \label{fig:3d-ffs-b}
\end{figure}

By repeating this analysis with multiple volumes, we consider the extrapolation to infinite volume.  Here we consider three different volumes, corresponding to five, six and seven-layers of cubes.  These correspond to 2643, 10497, and 41511 cubes, respectively.  Using the fit parameters from multiple volumes allows us to extrapolate to infinite cubes.  In Figs.~\ref{fig:3d-ffs-a} and ~\ref{fig:3d-ffs-b} we see the finite-size scaling of the fit parameters, $A$ and $B$, respectively, from Eq.~\eqref{eq:two-param-fit}.
The fit is of the form of Eq.~\eqref{eq:ffs-eqs}, with $N_{bound}$ being the size of the two-dimensional boundary of the three-dimensional hyperbolic lattice.
We find $A_{\infty} \simeq 1.11(2)$, and $B_{\infty} \simeq 1.62(9)$ in the infinite volume limit, with the squared boundary mass $M^{2} = 10$.

\section{Conclusions} \label{sec:conclusions}
In this paper we have studied the behavior of boundary correlations of massive scalar fields propagating on discrete tessellations of hyperbolic
space. Both two and three dimensions are examined and good quantitative agreement with the continuum formula relating the power of the boundary two-point correlator to the bulk mass is obtained. Specifically, the
functional form for the dependence of the boundary scaling dimension on bulk mass is reproduced accurately, including the inferred dimension
of the boundary theory. A single parameter, $B$,
identifies the effective squared radius of curvature of the lattice.  In fact, in two dimensions, if one considers the continuum formula for the squared radius of curvature for a $\{p,q\}$-tessellation of $\mathbf{H}^{2}$,
\begin{align}
\label{eq:roc}
    L^{2} = \frac{1}{4 \,  \arccosh^{2} \left( \frac{\cos(\pi/p)}{\sin(\pi/q)} \right) },
\end{align}
one would predict values of $0.84$, $0.43$ and $0.64$ for $\{3,7\}$, $\{3,8\}$ and $\{4,5\}$ tessellations, respectively.
These numbers differ from our fitted values for $B$ by a factor of approximately two (see the appendix).
This factor of two can be completely understood in terms of our choice of weights.  If one uses the dual weight prescription for the kinetic term as
given in \cite{brower2019lattice} which sets $p_{xy} V_{e} = \ell a$ where $\ell$ is the dual edge length, and $a$ is the lattice spacing, this factor of two arises naturally and gives the correct normalization for the squared radius of curvature.

In this work we held the lattice spacing fixed at $a=1$.  This appears in the results of the correlator at short distances as wiggles in the data, and at longer distances as a spread, and noise, in the correlators.  This is particularly apparent in the $\{4,5\}$ correlator data in the appendix.  One could refine the tessellation to approach the continuum manifold. In Ref.~\cite{brower2019lattice} they did exactly this and considered a refinement of the $\{3,7\}$ tesselation by continually inserting triangles inside of existing triangles at a fixed physical volume.

In the future we plan to extend these calculations to four dimensional hyperbolic space and to investigate the effects of allowing for dynamical fluctuations
in the discrete geometries in order to simulate the effect of gravitational fluctuations. In such scenarios,
the effects of the back reaction of matter fields on the geometries can be explored. This should allow us to probe holography in regimes
that are difficult to explore using analytical approaches.

\section*{Acknowledgements} \label{sec:acknowledgements}
SC and JUY would like to thank the QuLat collaboration and Rich Brower in particular for stimulating discussions.
This work is supported in part by the U.S.\ Department of Energy (DOE), Office of Science, Office of High Energy Physics, under Award Number {DE-SC0009998}
and {DE-SC0019139}.  Numerical computations were performed at Fermilab using USQCD resources funded by the DOE Office of Science.

\appendix
\section{Fits to $\{p,q\} = \{4,5\}$ and $\{p,q\} = \{3,8\}$}
\label{sec:45-fits}
\begin{figure}
    \centering
    \includegraphics[width=8.6cm]{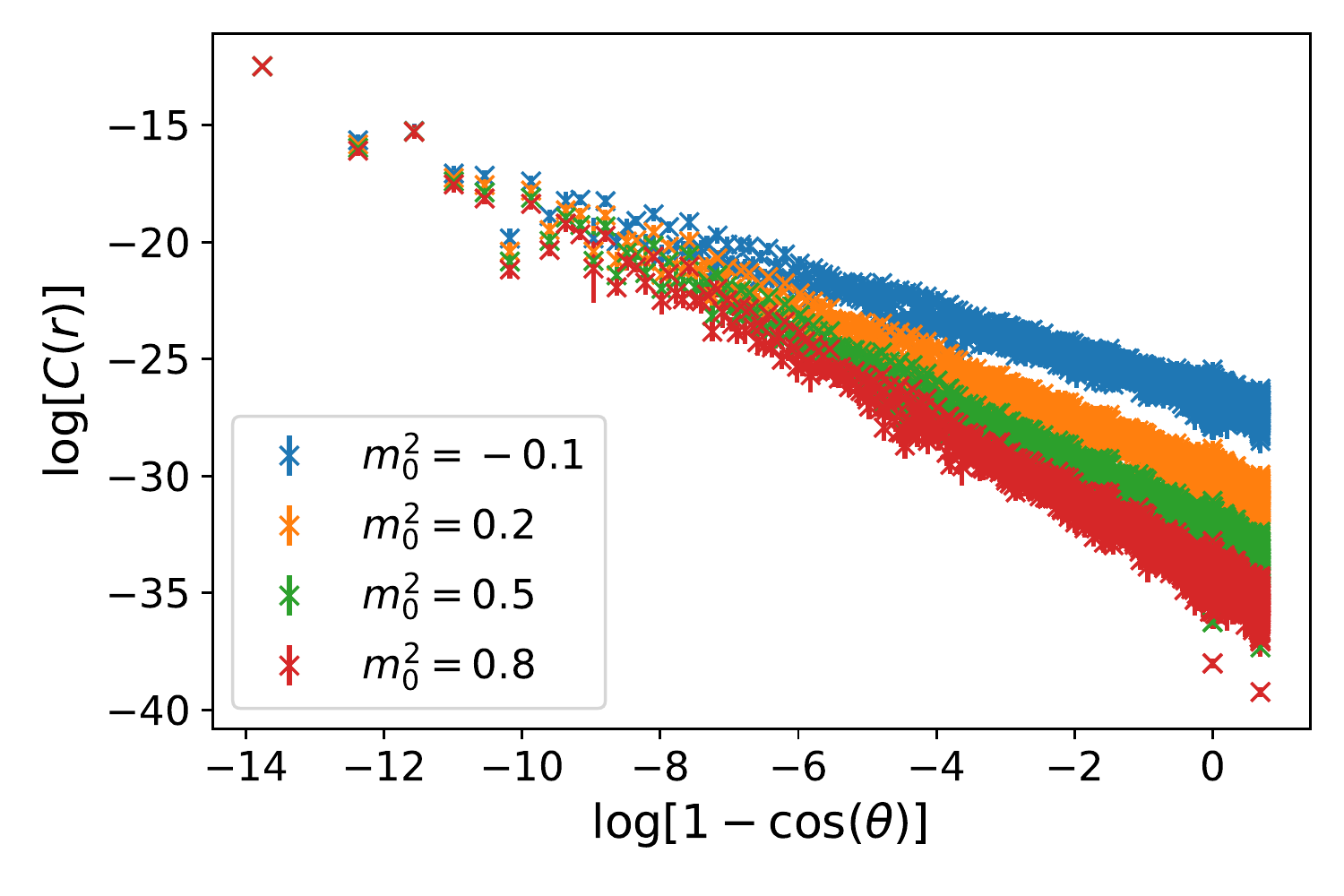}
    \caption{The boundary two-point correlator is plotted in log-log coordinates for the lattice $\{p,q\} = \{4,5\}$.  Here four bulk masses are shown for the values of $m_{0}^{2} = -0.1$, 0.2, 0.5, and 0.8, with a boundary mass of $M^2 = 500$.  The lattice is comprised of eight layers of squares.}
    \label{fig:45-corr}
\end{figure}
\begin{figure}
    \centering
    \includegraphics[width=8.6cm]{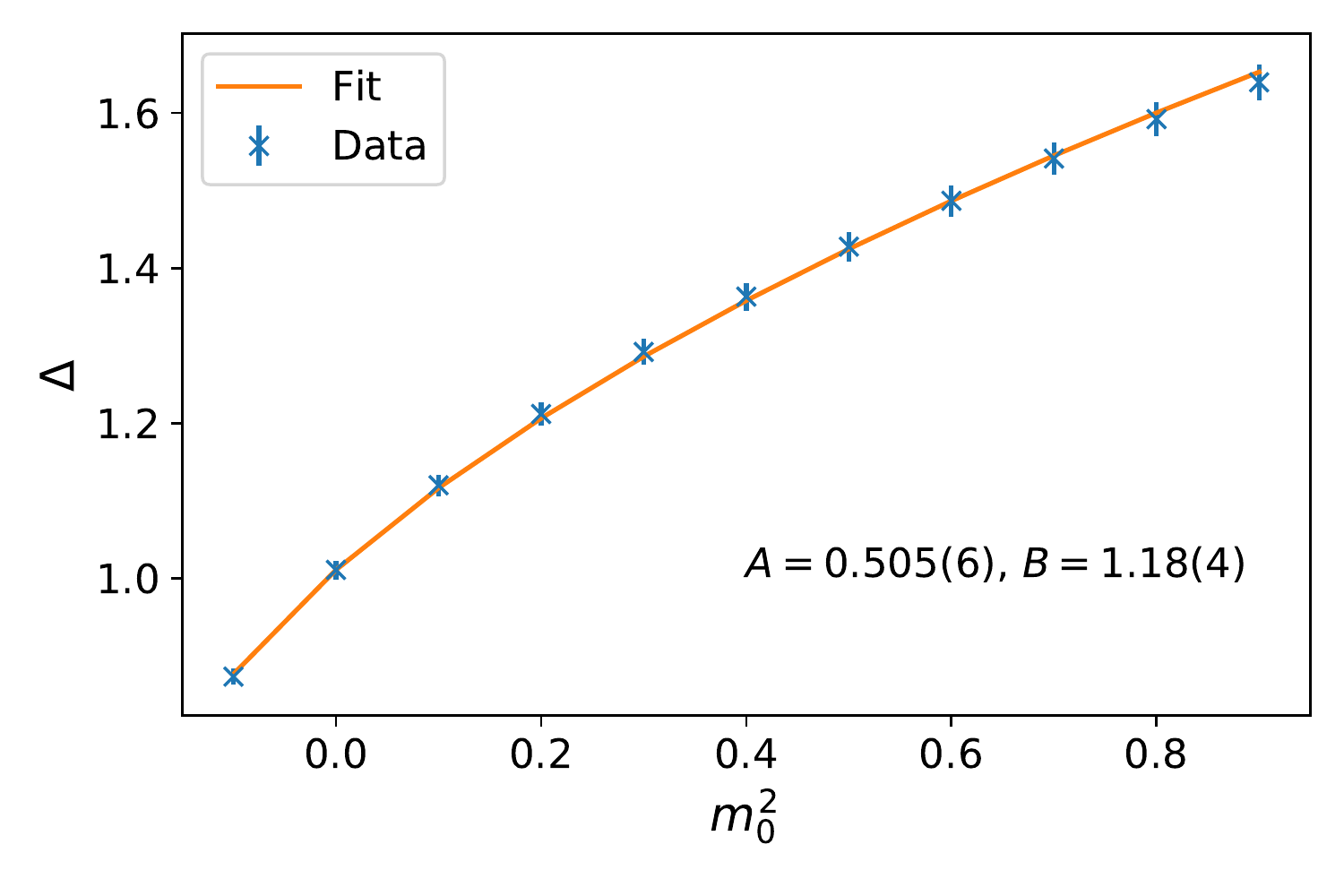}
    \caption{The power-law extracted from the correlator data for the $\{p,q\} = \{4,5\}$ lattice.  This data is for a eight layer lattice with a boundary mass of $M^2 = 500$.  The fit parameters for this fit are quoted inside the figure.}
    \label{fig:45-delta}
\end{figure}

\begin{figure}
    \centering
    \includegraphics[width=8.6cm]{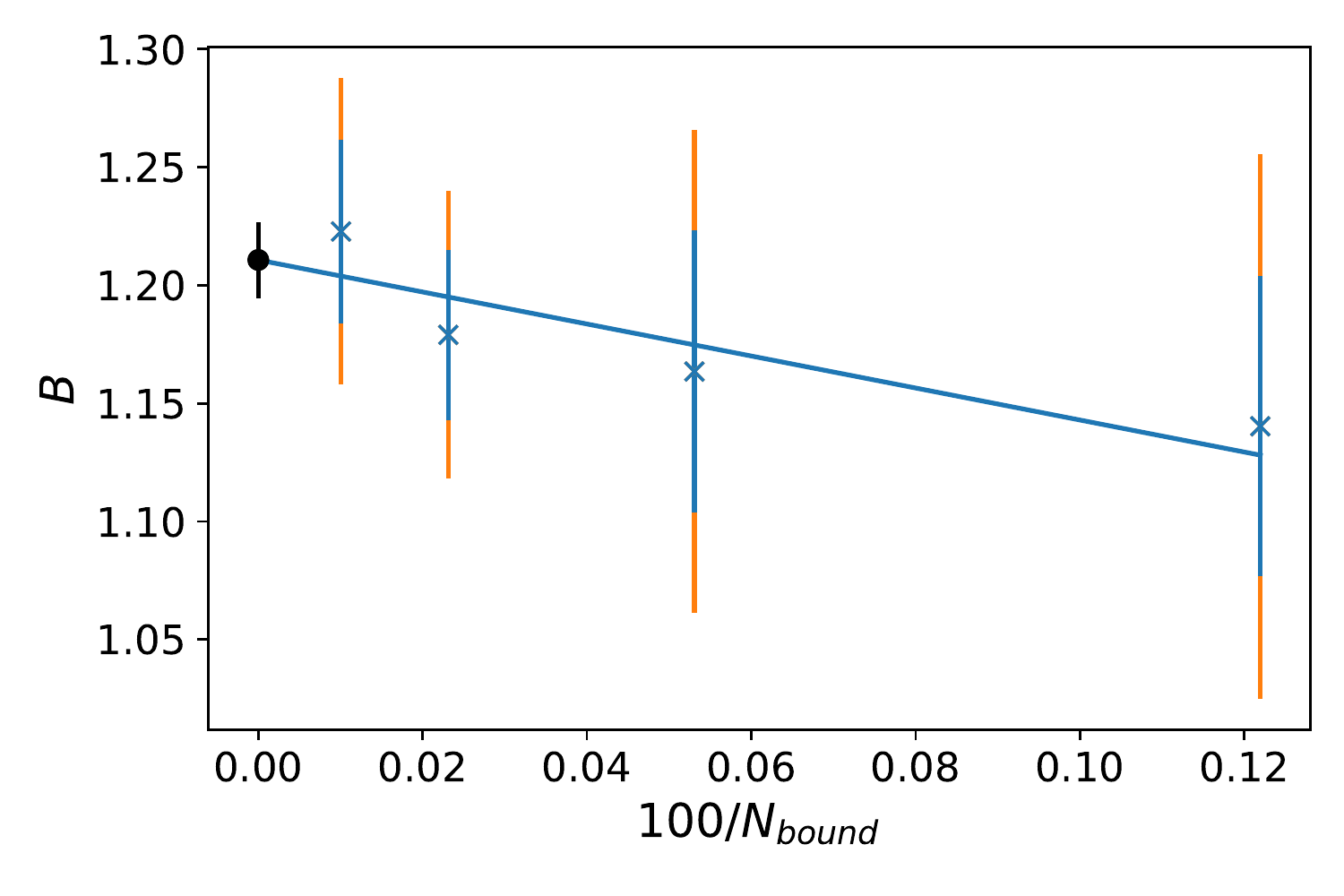}
    \caption{The finite-size scaling for the $B$ fit parameter versus the inverse boundary size.  Here we re-scaled the $x$-axis by 100 squares to declutter the axis labels.  This fit is for a boundary mass of $M^2 = 500$.}
    \label{fig:45-Bffs}
\end{figure}
\begin{figure}
    \centering
    \includegraphics[width=8.6cm]{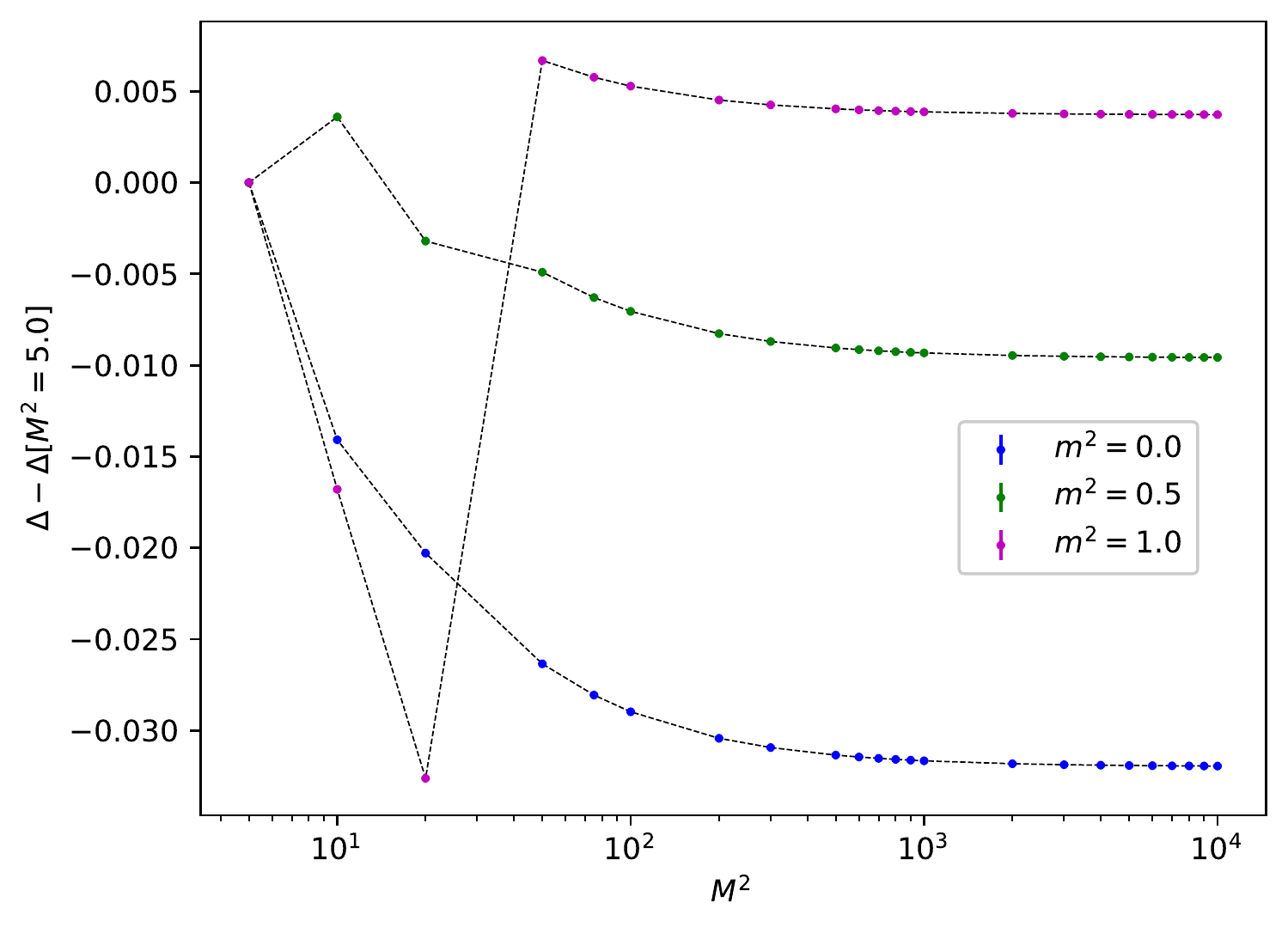}
    \caption{Boundary mass ($M^2$) dependence of the fit parameter $\Delta$ in the two point function for a 10 layer lattice in the $\{3,8\}$ geometry. Error bars are of the order $10^{-4}$ or smaller and not visible in the figure.}
    \label{boundaryMassDep}
\end{figure}
To ensure the fit parameters obtained in the $\{3,7\}$ study were not trivial, we also considered a $\{p,q\} = \{4,5\}$ and $\{p,q\} = \{3,8\}$  lattice, and repeated the analysis.  This should give the same result for $A$, since the boundary is still one-dimensional, however the curvature of the lattice is different from the $\{3,7\}$ case, and so we would expect a different result for the $B$ fit parameter.  In Fig.~\ref{fig:45-corr} we see an example of the boundary-to-boundary two-point correlator for four different masses for the $\{4,5\}$ tessellation.

There is more noise than in the $\{3,7\}$ case, perhaps due to the more coarse nature of the tessellation using squares instead of triangles; however, the power-law behavior is still apparent. We used binning in distance $r$ along boundary or equivalently in $\theta$ in the radial coordinates to reduce the fluctuation in the correlator data before performing the fits. This essentially reduces the fluctuation of the radial distance of the boundary points in the Poincar\'{e} disk picture while keeping the average radius of curvature unchanged, thus providing a better handle to predict the same continuum result.  
\begin{table*}[!hbtp]
\caption{\label{Tab1}}
\centering 
\makebox[\textwidth]{
\begin{tabular}{ |P{1cm}|P{1cm}|P{1cm}|P{2cm}|P{2cm}|P{2.5cm}|P{2cm}|P{2cm}|  }
 \hline
 \multicolumn{8}{|c|}{\{3,8\} tessellation results} \\
 \hline
 Layer & $r_{max}$ & bin-size & minimum fit-width &  maximum fit-width & fit-range & A & B \\
\hline
10 & 523 & 10 & 50 & 180 & 0$\leq \theta \leq$ 1.08 & 0.4926(4) & 0.757(3) \\
\hline
11 & 903 & 10 & 90 & 280 & $0\leq\theta\leq 0.97$ & 0.5000(6) & 0.776(5) \\
\hline
12 & 1554  & 15     &  150   & 495   &  $0\leq\theta\leq 1.00 $      & 0.4977(6)    & 0.82(1)         \\
 \hline
 \hline
- & $\infty$ & - &-&-&-& 0.503(5) & 0.81(2) \\
\hline
\hline
\end{tabular}}
\end{table*}

From the linear fit of the boundary-boundary correlators we extracted $\Delta$, which can be seen in Fig.~\ref{fig:45-delta}.
Again we see the boundary dimension parameter, $A$, gives a value similar to $d/2 = 1/2$.  By studying multiple volumes at sufficiently large boundary mass, $M^2 = 500$, we extrapolate to the infinite volume limit.  The result for the $B$ parameter can be seen in Fig.~\ref{fig:45-Bffs}. We find for our fits $A \simeq 0.506(2)$, and $B \simeq 1.21(2)$. 

We have performed a similar analysis for the $\{3,8\}$ case. In this case, the value of $\Delta$ settles down for boundary masses $M^2\ge 1000$ (Fig.~\ref{boundaryMassDep}). So $M^2=2000$ is picked to perform the analysis. The results are summarized in the table \ref{Tab1}. Notice that the fitted value for $B=0.81(2)$ is once more close to theoretical expectations which predict $L^2=B/2=0.43$ provided the kinetic term employs the dual lattice weight.  A summary of the fit ranges, bin sizes, and parameter values for the $\{3,8\}$ lattice can be found in table~\ref{Tab1}.

\bibliographystyle{unsrt}

\begin{thebibliography}{10}

\bibitem{Maldacena:1997re}
Juan~Martin Maldacena.
\newblock {The Large N limit of superconformal field theories and
  supergravity}.
\newblock {\em Int. J. Theor. Phys.}, 38:1113--1133, 1999.

\bibitem{Gubser:1998bc}
S.~S. Gubser, Igor~R. Klebanov, and Alexander~M. Polyakov.
\newblock {Gauge theory correlators from noncritical string theory}.
\newblock {\em Phys. Lett.}, B428:105--114, 1998.

\bibitem{Witten:1998qj}
Edward Witten.
\newblock {Anti-de Sitter space and holography}.
\newblock {\em Adv. Theor. Math. Phys.}, 2:253--291, 1998.

\bibitem{KLEBANOV199989}
Igor~R. Klebanov and Edward Witten.
\newblock Ads/cft correspondence and symmetry breaking.
\newblock {\em Nuclear Physics B}, 556(1):89 -- 114, 1999.

\bibitem{Krcmar_2008}
R~Krcmar, A~Gendiar, K~Ueda, and T~Nishino.
\newblock Ising model on a hyperbolic lattice studied by the corner transfer
  matrix renormalization group method.
\newblock {\em Journal of Physics A: Mathematical and Theoretical},
  41(12):125001, mar 2008.

\bibitem{PhysRevE.84.032103}
Seung~Ki Baek, Harri M\"akel\"a, Petter Minnhagen, and Beom~Jun Kim.
\newblock Ising model on a hyperbolic plane with a boundary.
\newblock {\em Phys. Rev. E}, 84:032103, Sep 2011.

\bibitem{Benedetti_2015}
Dario Benedetti.
\newblock Critical behavior in spherical and hyperbolic spaces.
\newblock {\em Journal of Statistical Mechanics: Theory and Experiment},
  2015(1):P01002, jan 2015.

\bibitem{PhysRevE.101.022124}
Nikolas~P. Breuckmann, Benedikt Placke, and Ananda Roy.
\newblock Critical properties of the ising model in hyperbolic space.
\newblock {\em Phys. Rev. E}, 101:022124, Feb 2020.

\bibitem{brower2019lattice}
Richard~C Brower, Cameron~V Cogburn, A~Liam Fitzpatrick, Dean Howarth, and
  Chung-I Tan.
\newblock Lattice setup for quantum field theory in ads $ \_2$.
\newblock {\em arXiv preprint arXiv:1912.07606}, 2019.

\bibitem{Zamolodchikov:2001dz}
A.~Zamolodchikov.
\newblock {Scaling Lee-Yang model on a sphere. 1. Partition function}.
\newblock {\em JHEP}, 07:029, 2002.

\bibitem{Brower:2018szu}
Richard~C. Brower, Michael Cheng, Evan~S. Weinberg, George~T. Fleming,
  Andrew~D. Gasbarro, Timothy~G. Raben, and Chung-I Tan.
\newblock {Lattice $\phi^4$ field theory on Riemann manifolds: Numerical tests
  for the 2-d Ising CFT on $\mathbb{S}^2$}.
\newblock {\em Phys. Rev. D}, 98(1):014502, 2018.

\bibitem{Coxeter1954}
Harold Stephen~Macdonald Coxeter.
\newblock Regular honeycombs in hyperbolic space.
\newblock In {\em Proceedings of the International Congress of Mathematicians
  of}, 1954.

\bibitem{Needham199902}
Tristan Needham.
\newblock {\em Visual Complex Analysis}.
\newblock Oxford University Press, USA, 1999.

\bibitem{nelson2017visualizing}
Roice Nelson and Henry Segerman.
\newblock Visualizing hyperbolic honeycombs.
\newblock {\em Journal of Mathematics and the Arts}, 11(1):4--39, 2017.

\end{thebibliography}

\end{document}